\documentclass[prb,aps,twocolumn,nopacs,superscriptaddress]{revtex4-2}
\usepackage[utf8]{inputenc}
\usepackage[T1]{fontenc}
\usepackage[english]{babel}
\usepackage[pdftex]{graphicx}  
\usepackage{graphicx, xcolor}
\usepackage{dcolumn}
\usepackage{bm}
\usepackage{amsmath,amsthm,amssymb}
\usepackage[T1,T2A]{fontenc}
\usepackage{xcolor}
\usepackage{graphicx}
\usepackage{braket}
\usepackage{wrapfig}
\usepackage{soul}
\usepackage{mathtools}
\usepackage{float}
\usepackage{tabularx}
\usepackage{makecell}
\usepackage{subfigure}
\usepackage{hyperref}
\hypersetup{colorlinks,bookmarksopen,bookmarksnumbered,
    citecolor=blue,
    linkcolor=blue,
    pdfstartview=false,
    urlcolor=blue}
\newcommand{\tb}{\textcolor{blue}}

\newcommand{\trace}{{\rm Tr}}

\begin{document}
\title{A Hybrid Anyon-Otto thermal machine}
\author{Mohit Lal Bera}
\affiliation{Departamento de Física Teórica and IFIC, Universitat de València-CSIC, 46100 Burjassot (València), Spain}
\affiliation{ICFO–Institut de Ciéncies Fotóniques, The Barcelona Institute of Science and Technology, Av. Carl Friedrich Gauss 3, 08860 Castelldefels (Barcelona), Spain}
\author{Joyce Kwan}
\affiliation{JILA, NIST, and Department of Physics, University of Colorado, Boulder, CO 80309, USA}
\author{Armando Pérez}
\affiliation{Departamento de Física Teórica and IFIC, Universitat de València-CSIC, 46100 Burjassot (València), Spain}
\author{Miguel A. García-March}
\affiliation{IUMPA - Instituto Universitario de Matemática Pura y Aplicada, Universitat Politècnica de València, E-46022 València,
Spain}
\author{Ravindra Chhajlany}
\affiliation{Institute of Spintronics and Quantum Information, Faculty of Physics and Astronomy,Adam Mickiewicz
University, 61614 Poznan, Poland}
\author{Tobias Grass}
\affiliation{DIPC - Donostia International Physics Center, Paseo Manuel de Lardizábal 4, 20018 San Sebastián, Spain}
\affiliation{IKERBASQUE, Basque Foundation for Science, Plaza Euskadi 5, 48009 Bilbao, Spain}
\author{Maciej Lewenstein}
\affiliation{ICFO–Institut de Ciéncies Fotóniques, The Barcelona Institute of Science and Technology, Av. Carl Friedrich Gauss 3, 08860 Castelldefels (Barcelona), Spain}
\affiliation{ICREA, Pg. Lluis Companys 23, 08010 Barcelona, Spain}
\author{Utso Bhattacharya}
\affiliation{Institute for Theoretical Physics, ETH Zurich, Zurich, Switzerland}
\author{Sourav Bhattacharjee}
\email{email: sourav.offc@gmail.com}
\affiliation{ICFO–Institut de Ciéncies Fotóniques, The Barcelona Institute of Science and Technology, Av. Carl Friedrich Gauss 3, 08860 Castelldefels (Barcelona), Spain}
\affiliation{Max Planck Institute for the Physics of Complex Systems, Nöthnitzer Str. 38, 01187 Dresden, Germany}

\maketitle
\section*{Abstract}
    We propose a four-stroke quantum thermal machine based on the 1D anyon Hubbard model, which is capable of extracting the excess energy arising from anyon exclusion statistics at low temperature into finite work. Defining a hybrid anyon-Otto (HAO) cycle, we find that the low-temperature work, in the absence of any interactions, is maximized in the pseudo-fermionic limit, where the anyons most closely resemble free fermions. However, when weak interactions are introduced, the work output is no longer maximized at the bosonic or pseudo-fermionic extremes but instead peaks at intermediate statistical angles. This clearly demonstrates that interactions and anyonic statistics conspire non-trivially to enhance performance, with interacting anyons offering greater quantum thermodynamic advantage than either bosons or pseudo-fermions, in this regime. Furthermore, we also outline an experimental protocol to realize the HAO cycle using ultracold atoms in an optical lattice.  
\section*{Introduction}\label{sec:intro}
In recent times, a plethora of models of quantum thermal machines (QTMs) have been proposed and analyzed to understand the emergence of thermodynamic principles from quantum dynamics \cite{kosloff14, souravReview21}. The existence, albeit mostly theoretical, of such a large number of models can be attributed to two reasons. Firstly, the nondeterministic and destructive nature of generic quantum measurements makes it difficult to uniquely and unambiguously define 
the quantum equivalents of classical work and heat. Secondly, genuine quantum phenomena have been shown to alter the performance of QTMs in a myriad of ways, depending on how they are incorporated into the model, for example, in the form of coherence in the system \cite{hardal15, camati19, mohan25}, engineered heat baths \cite{scully03, lutz14, manzano16, klaers17, Niedenzu18, ray23, bera24, sen25}, many-body effects \cite{campisi16critical, fazio16,hartmann20, revathy20, piccitto22, solfanelli23, arezzo24, mukherjee24, brollo25}, etc. In addition to theoretical investigations, progress has also been made in the experimental realization of QTMs \cite{von19, gluza21, Koch2023, Aamir2025, uusnäkki25experiment}.

\begin{figure}[t]
    \centering
    \includegraphics[width=0.98\linewidth]{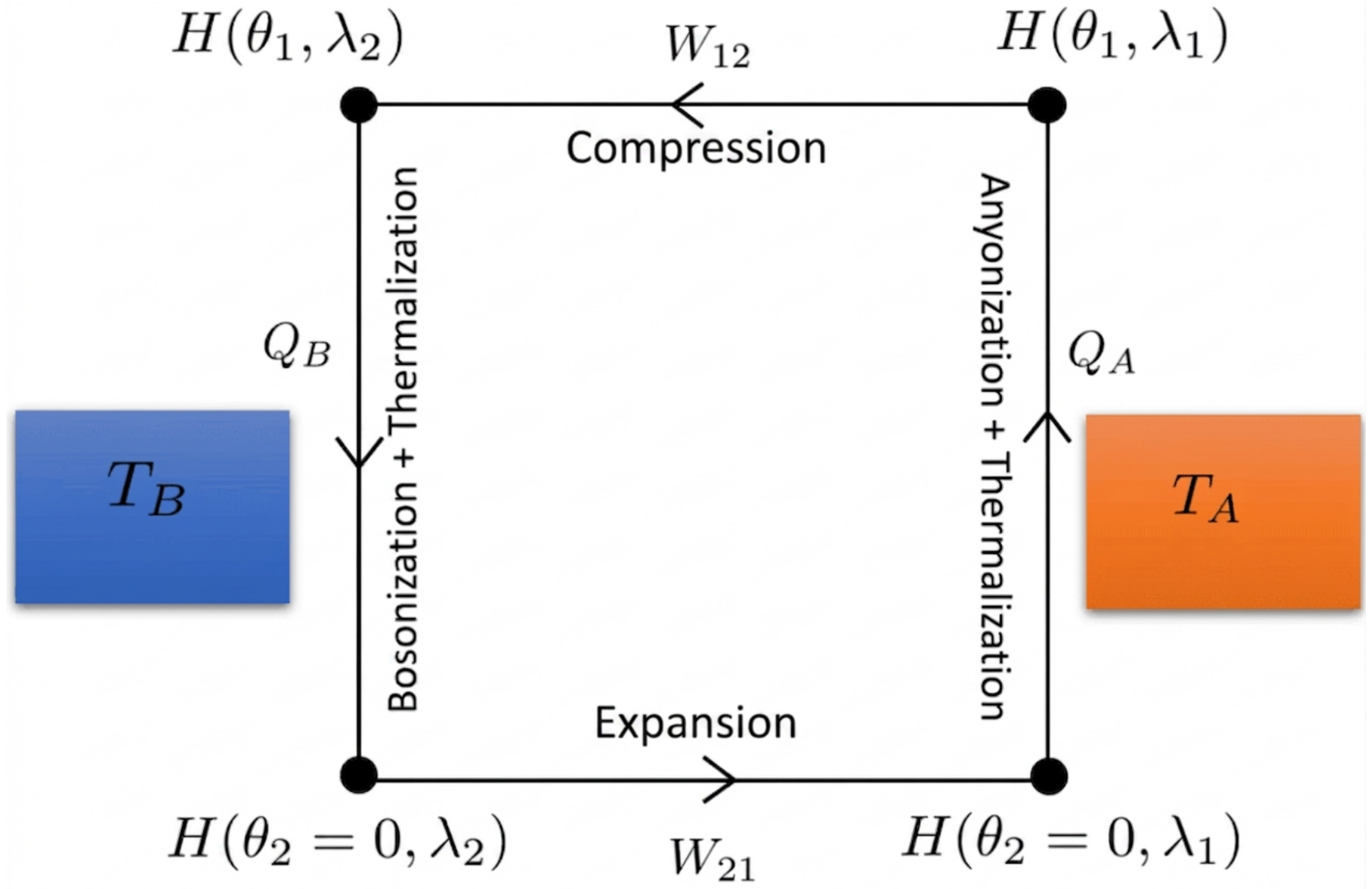}
    \caption{{\bf Schematic representation of the HAO cycle}. The two unitary work strokes - expansion and compression, are implemented through an explicit change of the Hamiltonian parameter $\lambda\in\{J,U\}$. The heat strokes consist of thermalization with one of the heat baths with temperature $T_A$ or $T_B$ and a change in the statistical parameter $\theta$.}
    \label{fig:schematic}
\end{figure}
\begin{table*}[t]
    \centering
    \begin{ruledtabular}
    \begin{tabular}{cccc}
          Mode & \makecell{Total work extracted} & \makecell{Energy gained from hotter bath} & \makecell{Energy gained from colder bath} \\
          \hline
          Engine (E) & $+$ & $+$ & $-$ \\
          Refrigerator(R) & $-$ & $-$ & $+$ \\
          Accelerator (A) & $-$ & $+$ & $-$ \\
          Inverse Accelerator (IA) & $+$ & $-$ & $+$ \\
          Heater (H) & $-$ & $-$ & $-$ 
    \end{tabular}
    \end{ruledtabular}
    \caption{{Summary of the different modes of operation of the HAO cycle. The $+(-)$ symbol in the second column corresponds to a net work extracted from (performed on) the system. Similarly, in the third and fourth column, $+(-)$ corresponds to energy gained (lost) by the system from the respective baths.}}
    \label{table:summary}
\end{table*}

The simplest and most commonly studied models of QTMs are those based on the four-stroke quantum Otto cycle \cite{souravReview21}. The standard Otto cycle consists of two \textit{ work strokes} and two \textit{thermalization strokes}. In the former, a system is made to evolve unitarily through a quench in some parameter of the Hamiltonian, with the energy change of the system associated with work. The work strokes are interceded by the thermalization strokes in which the system evolves dissipatively in contact with a \textit{heat bath}, the energy change being associated with the heat transferred. At the end of the full cycle, the system returns to its initial state with a net conversion of quantum heat into quantum work or vice versa. The direction of heat flow and work output determines the operating mode of the cycle, with the thermodynamically allowed (respecting the second law) modes being the engine, refrigerator, accelerator, and heater. The engine (refrigerator) mode is characterized by a net heat transfer from the hotter (colder) to the colder (hotter) bath, with a net work output (input). In accelerator mode, input work is used to boost the heat transfer from the hotter to the colder bath, while in heater mode, input work is fully converted into heat and dumped into each of the baths.

\begin{figure*}
    \centering
    \includegraphics[width=\linewidth]{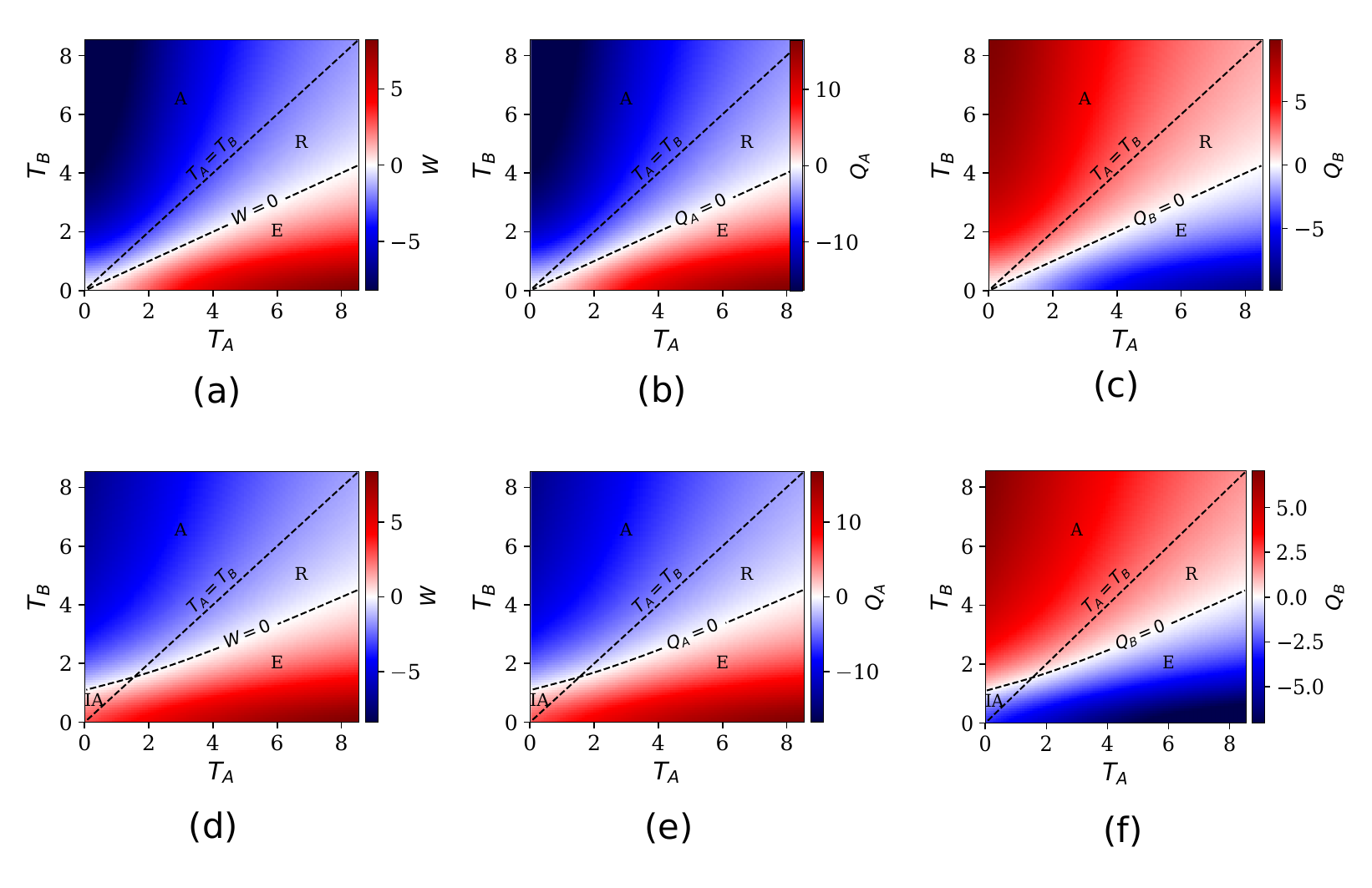}
    \caption{{\bf Operating modes of the HAO cycle in the non-interacting limit.} Work output $W$ ((a) and (d)), heat exchanged with the baths $Q_A$ ((b) and (e)), and $Q_B$ ((c) and (f)), in the HAO cycle as a function of the bath temperatures $T_A$ and $T_B$. The top panel corresponds to the standard Otto cycle with $\theta_1=\theta_2=0$, showing the three possible working modes - engine(E), refrigerator (R) and accelerator (A). The bottom panel shows the emergence of the inverse accelerator mode (IA) in the low temperature limit for $\theta_2=0, \theta_1\neq 0$. The plots depicted above are obtained numerically for a chain of length {$L=12$, with number of particles $N=6$, $U=0$, $J_1=2.0$ and $J_2=1.0$}.}
    \label{fig:modes_noninter}
\end{figure*}

The Otto cycle operating in the engine mode, by definition, relies on thermal energy for work output; the latter thus vanishes in the limit of zero temperature of the heat baths. However, quantum statistical properties depend crucially on the nature of particles at low temperatures, and this has recently been exploited to design and experimentally realize the Pauli engine~\cite{Koch2023}. In this variant of the regular Otto engine, the thermalization strokes are replaced by the so-called Pauli strokes, in which the system undergoes a bosonization or fermionization process. Specifically, rather than thermalizing with a heat bath, the statistical nature of the system is altered by tuning it from a molecular Bose condensate to a Fermi gas or vice versa. By virtue of the Pauli exclusion principle, the system has a higher energy ground state in the fermionic state as compared to the bosonic state. The Pauli engine operates by extracting this difference in ground state energy, defined as the Pauli energy, into useful work. Importantly, the Pauli engine has no classical analogue and is purely driven by quantum statistical phenomena. To this end, we note that the relation between the performance of QTMs and the statistical properties of the system has been explored in a number of other works, especially with respect to fermionic and bosonic statistics \cite{myers20, watanabe20, souza22, myers22, ghosh23,Boubakour2023, Menon2025}. 

In parallel, it has also been shown in recent times that exotic quantum particles such as anyons \cite{Kundu99,anyonRev08,sternRev08, Iqbal24, Zhang25, Himanshu25, Ghosh25, Yutushui25}, including impurity excitations in fractional quantum Hall liquids which exhibit fractional angular momentum and effective anyonic statistics \cite{grass20,baldelli21}, and paraparticles \cite{hazzard25,Mostaan25}, satisfy statistical properties different from bosons or fermions.
The performance of QTMs based on systems satisfying such non-trivial statistics has largely remained unexplored, except recent models based on few-body anyonic systems \cite{Myers2021,manianyonotto24, manianyoncarnot25, Dunlop25}.
In this work, we establish how anyonic statistics leads to non-trivial performance enhancement in a many-body QTM which, importantly, can be realised using current experimental platforms. Specifically, we define and analyze a hybrid Anyon-Otto (HAO) cycle (see Fig.\ref{fig:schematic}), based on the anyon Hubbard model (AHM) \cite{keilmann11, Longhi12, anyonhubbard15,Tang15,ArcilaForero16,Zhang17, Zhang20,Bartolomei20,Bonkhoff21,Bonkhoff23, greinerExpt24, Dhar2025, Wang25, Bonkhoff2025, Theel25}, in which the statistical properties of the particles can be continuously tuned from boson-like to pseudo-fermion-like (satisfy anti-commutation relations only when particles are on different sites). We contrast the operation of the HAO cycle with the Otto cycle in the same temperature and parameter regime except that the statistical properties of the system remain unchanged throughout the Otto cycle. We find that the HAO cycle can produce a finite work output even in the limit of vanishing temperatures of the baths, due to the presence of an \textit{anyon energy}, defined analogously to the Pauli energy. In the absence of explicit interactions between the anyons, we demonstrate that the HAO cycle can operate in an \textit{inverse accelerator} mode for finite but small temperatures. This mode, characterized by a heat transfer from colder to hotter bath and a net work output, exhibits an apparent violation of the second law which is reconciled when the work production in the anyonization stroke is appropriately accounted for. In presence of weak interactions between the anyons, we observe that anyonic statistics boosts the performance of the QTM at low temperatures, manifested in the form of higher work output as compared to bosonic or pseudo-fermionic statistics. Finally, we propose how the HAO cycle can be realized experimentally through a feasible extension of the experimental protocol used for the recent realization of the AHM \cite{keilmann11,greinerExpt24, Dhar2025}, thus rendering the possibility of direct experimental verification of our results.

\section*{Results}

\paragraph*{\bf Anyon Hubbard model}
The 1D AHM is realized by introducing a synthetic gauge field in the hopping term of the Bose Hubbard model \cite{anyonhubbard15},
\begin{equation}\label{eq:ham}
    H = -J\sum_{j=1}^{L}\left(\hat{b}_j^\dagger e^{-i\hat{n}_j\theta}\hat{b}_{j+1} + h.c.\right) + \frac{U}{2}\sum_{j=1}^{L} \hat{n}_j\left(\hat{n}_j-1\right),
\end{equation}
where $L$ is the total number of sites, $\hat{b}_j$ are the bosonic annihilation operators, $\hat{n}_j=\hat{b}_j^\dagger \hat{b}_j$ is the number operator,  $J$ is the tunneling amplitude, $U$ is the interaction strength, and $\theta$ is a phase parameter. Defining the annihilation operators $\hat{a}_j = e^{-i\theta\sum_{1\leq l\leq j-1}\hat{n}_j}\hat{b}_j$, the equivalent anyon model is obtained,
\begin{equation}\label{eq:hamAnyon}
    H = -J\sum_{j=1}^{L}\left(\hat{a}_j^\dagger \hat{a}_{j+1} + h.c.\right) + \frac{U}{2}\sum_{j=1}^{L} \hat{n}_j\left(\hat{n}_j-1\right),
\end{equation}
where $\hat{n}_j = \hat{b}_j^\dagger \hat{b}_j = \hat{a}_j^\dagger \hat{a}_j$.
The operators $\hat{a}_j$ satisfy the anyonic commutation relations, $\hat{a}_j\hat{a}_k^\dagger - e^{i\theta{\rm sgn}(j-k)}\hat{a}_k^\dagger \hat{a}_j=\delta_{jk}$ and $\hat{a}_j\hat{a}_k - e^{i\theta{\rm sgn}(j-k)}\hat{a}_k \hat{a}_j=0$. The phase parameter $\theta$ thus provides a direct control over the statistical properties of the particles. It is straightforward to see that the bosonic statistics are recovered in the limit $\theta\to 0$. In contrast, $\theta\to\pi$ leads to pseudo-fermionic statistics as the anyon operators anti-commute  on different lattice sites but commute on the same site.

\begin{figure}
    \centering
    \includegraphics[width=\linewidth]{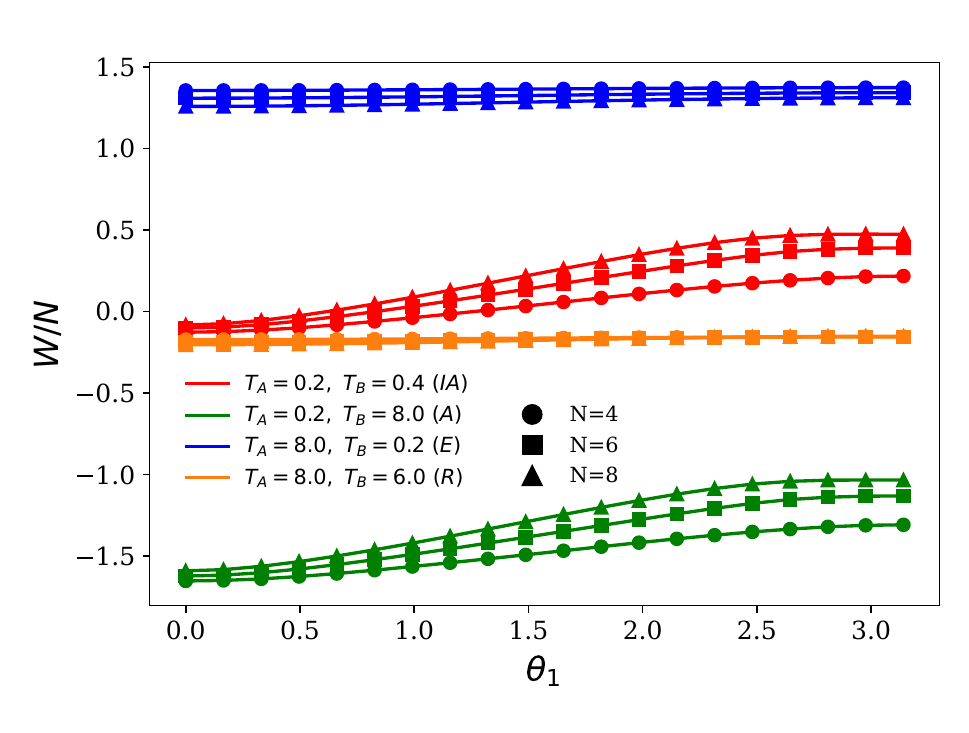}
    \caption{{\bf Work output in the non-interacting limit $\mathbf{U=0}$.} The work output per particle in the non-interacting limit $U=0$ is shown as a function of $\theta_1$. The anyon energy which is built up during the anyonization stroke is finite only when $T_A\to 0$. The work output due to the anyon energy increases monotonically with $\theta_1$. The plots are obtained numerically for $L=12$, $\theta_2=0$, $J_1=2.0$ and $J_2=1.0$.}
    \label{fig:work_theta}
\end{figure}
\begin{figure*}
    \centering
    \includegraphics[width=\linewidth]{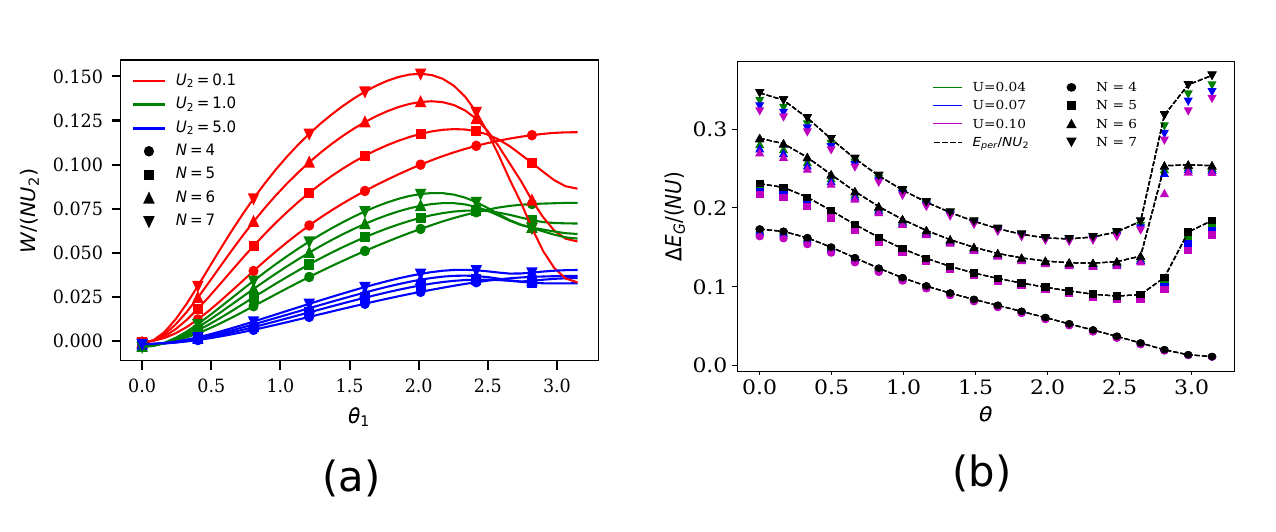}
    \caption{{\bf Work output in the interacting limit $\mathbf{U\neq 0}$}. (a) Low temperature ($T_A=T_B=0.1$) work output per particle (scaled with $U_2$) in the presence of finite interaction $U_2\neq0$ as a function of $\theta_1$. On approaching the weakly interacting limit $U_2\ll J$, the  work output exhibits a non-monotonic behavior with $\theta_1$ {as $N\to \frac{L}{2}^-$ }. The plots are obtained numerically for {$L=12$}, $J=1.0$, $U_1=0$ and $\theta_2=0$. (b) Increase in the ground state energy $\Delta E_G$ when interaction is switched on to a finite $U$. The increase is well approximated (black dashed lines) by first order correction to the ground state energy with $U$ considered as a small perturbation.}
    \label{fig:work_finiteU}
\end{figure*}
\paragraph*{\bf Hybrid Anyon-Otto cycle}\label{subsec:pauliOtto}
The HAO cycle that we propose here is a  four-stroke cycle that inherits operational characteristics from both the Pauli and Otto cycles. Thus, a hallmark of this cycle is its ability to function both as an anyon engine (defined analogously to the Pauli engine) at low temperature and as the standard Otto engine at high temperature. Before defining the HAO cycle, we recall that in the Otto cycle, work and heat are commonly associated with the energy exchange in strokes undergoing exclusively unitary or dissipative dynamics, respectively,
\begin{equation}
W_{otto}=\int\trace\left[\rho\frac{\partial H}{\partial t}\right]dt,
\end{equation}
\begin{equation}
Q_{otto}=\int\trace\left[\frac{\partial \rho}{\partial t}H\right]dt,
\end{equation}
where $\rho$ is the density operator representing the state of the system. In other words, a unitary evolution with a time-dependent Hamiltonian only contributes to the work, while a dissipative evolution, such as thermalization with a heat bath, only results in heat transfer in the Otto cycle. 

We now define the HAO cycle as follows. At the start of the cycle, we assume that the system with Hamiltonian $H(\theta_1, \lambda_1)$ is in equilibrium with a thermal bath with temperature $T_A$ (inverse temperature $\beta_A$), and $\lambda$ representing either of the parameters $J$ or $U$ appearing in the Hamiltonian. The initial state is therefore a thermal Gibbs state $\rho_1\equiv e^{-\beta_A H(\theta_1, \lambda_1)}/ \trace\left[{e^{-\beta_A H(\theta_1, \lambda_1)}}\right]$. The HAO cycle then consists of the following sequential strokes (see Fig.~\ref{fig:schematic}):

\begin{enumerate}
    \item \textit{Unitary {compression}}: ($\lambda_1 \to \lambda_2$, isolated) - The Hamiltonian parameter is ramped from $\lambda_1$ to $\lambda_2$ over a time $\tau$. The system evolves unitarily to a state $\rho_2 = U_{12}^\dagger(\tau)\rho_1 U_{12}(\tau)$ during this time interval. The change in the energy expectation value is thus associated with work performed,
    \begin{equation}\label{eq_W12}
        W_{12} = -\Big[\trace\left(\rho_2  H(\theta_1, \lambda_2) \right) - \trace\left(\rho_1 H(\theta_1, \lambda_1)\right)\Big].
    \end{equation}
    \item \textit{Thermalization B}: ($\theta_1\to\theta_2$, contact with heat bath) - This stroke consists of two sub-strokes -- the phase parameter is tuned from $\theta_1$ to $\theta_2$ followed by thermalization with a heat bath of temperature $T_B$. Note that the two sub-strokes can be carried out simultaneously, provided that the phase parameter is changed over a time interval much shorter than the thermalization time-scale. The steady state reached is thus given by, $\rho_3 = e^{-\beta_B H(\theta_2, \lambda_2)}/ \trace\left[{e^{-\beta_B H(\theta_2, \lambda_2)}}\right]$. The change in the energy expectation value in this stroke has two contributions, one from the energy dissipated to the heat bath and the other from the work done in changing the phase parameter. However, for reasons that we shall clarify below, we associate the total energy change in this stroke with heat transfer,
    \begin{equation}
        Q_B = \Big[\trace\left(\rho_3 H(\theta_2, \lambda_2)\right) - \trace\left(\rho_2 H(\theta_1, \lambda_2) \right)\Big].
    \end{equation}
    \item \textit{Unitary {expansion}}: ($\lambda_2 \to \lambda_1$, isolated) - As in the case of the unitary {compression} stroke, the system evolves unitarily to $\rho_4 = U_{21}^\dagger(\tau)\rho_3 U_{21}(\tau)$ and the change in the energy expectation value corresponds to work performed $W_{21}$,
    \begin{equation}\label{eq_W21}
    W_{21} = -\Big[\trace\left(\rho_4  H(\theta_2, \lambda_1) \right)- \trace\left(\rho_3 H(\theta_2, \lambda_2)\right)\Big].
    \end{equation}
    \item \textit{Thermalization A}: ($\theta_2\to\theta_1$, contact with heat bath) - In the final stroke, the phase parameter is restored to its initial value $\theta_1$ and the subsequent thermalization with heat bath at temperature $T_A$ results in the system returning back to its initial state $\rho_1$. Thus,
    \begin{equation}
         Q_A = \Big[\trace\left(\rho_1 H(\theta_1, \lambda_1)\right) - \trace\left(\rho_4 H(\theta_2, \lambda_1) \right)\Big].
    \end{equation}
\end{enumerate}
The total work $W=W_{12}+W_{21}$, $Q_A$ and $Q_B$, satisfies energy conservation (first law) $W=Q_A+Q_B$. The difference of the HAO cycle defined above with the standard quantum Otto cycle lies in the two modified thermalization strokes. In the latter case, the statistical properties of the quantum model remain unchanged, and thus the heat change is solely associated with the energy gained or lost to the heat baths. In contrast, thermalization in the HAO cycle is accompanied by a explicit change of statistical properties through the parameter $\theta$. It is important to note that ramping the phase parameter modifies the Hamiltonian itself and thus technically amounts to performing a certain amount of work. However, in the anyon equivalent of the model defined in Eq.~\eqref{eq:hamAnyon}, $\theta$ implicitly controls the statistical properties and hence the equilibrium configuration of the system. Since the purpose of the HAO engine is to convert both thermal energy and anyon energy into useful work, we associate the total energy change in the modified thermalization strokes only with heat energy.

\begin{figure}
    \centering
    \includegraphics[width=\linewidth]{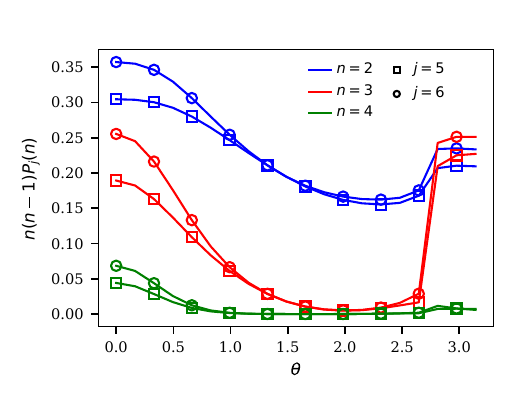}
    \caption{{\bf Particle occupancy probabilities in the ground state for $\mathbf{U=0}$}. The probability $P_j(n)$ of $n$ bosons (scaled with $n(n-1)$) occupying the $j^{th}$ site in the chain in the ground state is shown as a function of $\theta$. The plots are obtained for a system of size $L=12$ with $N=6$, $J=1.0$ and $U=0$.}
    \label{fig:occupancy_theta}
\end{figure}

\paragraph*{\bf The non-interacting limit $\mathbf {U=0}$}

We first consider the operation of the HAO cycle in the non-interacting limit $U=0$ with $\lambda\equiv J$. In this limit, the Hamiltonians at different instants of time during the work strokes commute with each other, implying adiabatic evolution and thus rendering the cycle operation  independent of the work stroke duration $\tau$.  We also assume that the total number $N$ of particles remain conserved throughout the cycle.  In Fig.~\ref{fig:modes_noninter} \tb{(a)-(c)}, we examine the possible modes of operation for the standard Otto cycle ($\theta_1 = \theta_2 = 0$), choosing without loss of generality $J_1=2.0$, $J_2 = 1.0$, {$L=12$ and $N=6$}. Depending on the signs of $W$, $Q_A$ and $Q_B$, and the relative magnitudes of $T_A$ and $T_B$, three modes of cycle operation can be identified. {When $T_A>T_B$, the cycle can operate either in the engine mode (E) with $Q_A, W >0$, $Q_B<0$ or the refrigerator mode (R) with $Q_A, W <0$, $Q_B>0$. For $T_A<T_B$, the only possible mode of operation is the accelerator mode (A) with $Q_A, W <0$, $Q_B>0$ (see Table.~\ref{table:summary})}. The E-R transition can be identified by the line $W = Q_A = Q_B = 0$. For $T_A, T_B  \to 0$, the work output and the heat vanish identically.

Having established the modes of operation in the Otto cycle, we now consider the situation $\theta_1 = \pi$,  $\theta_2=0$, that is, the statistical properties of the system are altered between the {pseudo-fermionic and bosonic limits} during the thermalization strokes. From Fig.~\ref{fig:modes_noninter} \tb{(d)-(f)}, it is evident that in the large temperature limit of the baths, $T_A, T_B \gg 1$, the HAO cycle is identical to the Otto cycle owing to the fact that the anyonic statistics can be well-approximated by Boltzmann statistics in this limit. However, at low temperatures, $T_A, T_B \to 0$, the excess energy resulting from the development of anyonic exclusions begins to manifest itself in the form of the non-zero work output and heat exchanges. This leads to the emergence of an \textit{inverse accelerator} (IA) mode - heat flows from the colder bath to hotter bath with a net 
work extraction, $Q_A, W >0$, $Q_B<0$ for $T_A<T_B$. As we shall show later, the IA mode does not violate the second law of thermodynamics since the anyon energy gained by the system during the anyonization process involves an additional work cost, which when included in the definition of $W$ restores the second law.

The finite work output in the limit of vanishing temperature manifests itself not only in the pseudo-fermionic limit $\theta_1=\pi$, but for any $\theta_1\neq 0$. This is demonstrated in Fig.~\ref{fig:work_theta}, where we observe that the average work output per particle $W/N$ increases  monotonically with increasing $\theta_1$ in the inverse accelerator mode ($T_A=0.2$, $T_B=0.4$), provided $T_A\to 0$. Similarly, the work input decreases with increasing $\theta_1$ in the accelerator mode {($T_A=0.2$, $T_B=8.0$)} in the same limit of $T_A$. Note that it is sufficient for only $T_A$ to be small as the anyonization process occurs in contact with the heat bath of temperature $T_A$.

\paragraph*{\bf Weak interaction $\mathbf{U\ll J} $}\label{subsec:weakU}
We now consider the situation in which work is performed by ramping the interaction parameter $U$ during the work strokes, holding $J$ to a constant value. 
In this case, the Hamiltonian does not commute with itself at different times during the work strokes, which leads to non-adiabatic excitations for finite stroke duration $\tau$. However, for simplicity, we shall assume that  $\tau\to\infty$ as the results discussed in the following do not change qualitatively in the presence of non-adiabatic excitations. In order to ensure that the anyon energy is not affected by the interaction, we choose $U_1 = 0$. The work output $W$ at low temperature, $T_A=T_B=0.1$, as a function of the statistical parameter $\theta_1$ is shown in Fig.~\ref{fig:work_finiteU}\tb{(a)} for different interaction strengths $U_2$. In the weakly interacting limit $U\ll J$, we observe that $W$ is maximized for $\theta_1 = \theta^*$ such that $0<\theta^*<\pi$, for $N\geq L/2$. However, as the interaction strength is  increased, the maxima shifts towards the pseudo-fermionic limit $\theta^*=\pi$. {Furthermore, within the largest Hilbert-space dimensions numerically accessible to us, we do not observe any appreciable drift of $\theta^*$ with increasing $L$ as long as the filling fraction $N/L$ is held constant \cite{supp}. This indicates that an intermediate optimal value of $\theta$ is likely to persist in the thermodynamic limit.}

To understand the above results, we first note that the dependence of $W$ on $\theta_1$ arises only from the unitary {compression} stroke, that is, when the interaction strength is increased from $U_1=0$ to $U_2$ at a fixed $\theta_1$.  In Fig.~\ref{fig:work_finiteU}\tb{(b)}, we see that the change in the ground state energy $\Delta E_G$ during this stroke is also minimized at intermediate values of $\theta_1$ between $0$ and $\pi$ for $U\ll J$ and $N\geq L/2$. The maximization of $W$ (recall that it is defined with a negative sign) observed in Fig.~\ref{fig:work_finiteU}\tb{(a)} is a direct consequence of this non-trivial behavior of the ground state energy. {This can be understood as follows. Firstly, for $U\ll J$, $\Delta E_G$ can be approximated by the first order perturbative correction to the ground state energy of the non-interacting part of the Hamiltonian. Considering the interaction $H_{int}=\frac{U}{2}\sum_j \hat{n}_j\left(\hat{n}_j-1\right)$ as a perturbation of magnitude $U$,
\begin{equation}\label{eq_Eper}
    \Delta E_G\approx E_{per} \!=\! \braket{\psi_0|H_{int}|\psi_0}\!=\!\frac{U}{2}\!\sum_j^L\!\left[\!\sum_{n=0}^N\! n(n-1)P_j(n)\right],
\end{equation}
where $\ket{\psi_0}$ is the ground state of the system for $U=0$, and $P_j(n)$ is the marginal probability that the $j^{th}$ site is occupied by $n$ particles in the ground state $\ket{\psi_0}$ (see \cite{supp} for detail). Figure~\ref{fig:work_finiteU}\tb{(b)} shows that $E_{per}$ (dashed lines) indeed approximates $\Delta E_G$ for $U\ll J$. Secondly, from Eq.~\eqref{eq_Eper}, it is clear that $E_{per}$ depends only on the probabilities of double or higher occupancies per site. The relatively smaller magnitude of $E_{per}$ for intermediate values of $\theta$ for $N\geq L/2$ can therefore be attributed to smaller values of $P_j(n\geq2)$ at these values as compared to $\theta=0$ and $\theta=\pi$.  To support the above argument, we plot in Fig.~\ref{fig:occupancy_theta} the marginal probability $P_j(n)$ that the $j^{th}$ bulk site (avoiding edge effects due to OBC) is occupied by $n_j$ particles for a couple of neighboring bulk sites {$j=5,6$ in the system with $L=12$}, $J=1.0$ and $U=0$. It can be seen from Fig.~\ref{fig:occupancy_theta} that the $P_j(n\geq 2)$ decreases initially as $\theta$ is increased from zero, before increasing again on approaching $\theta=\pi$. }

\begin{figure*}
    \centering
    
    \includegraphics[width=0.8\linewidth]{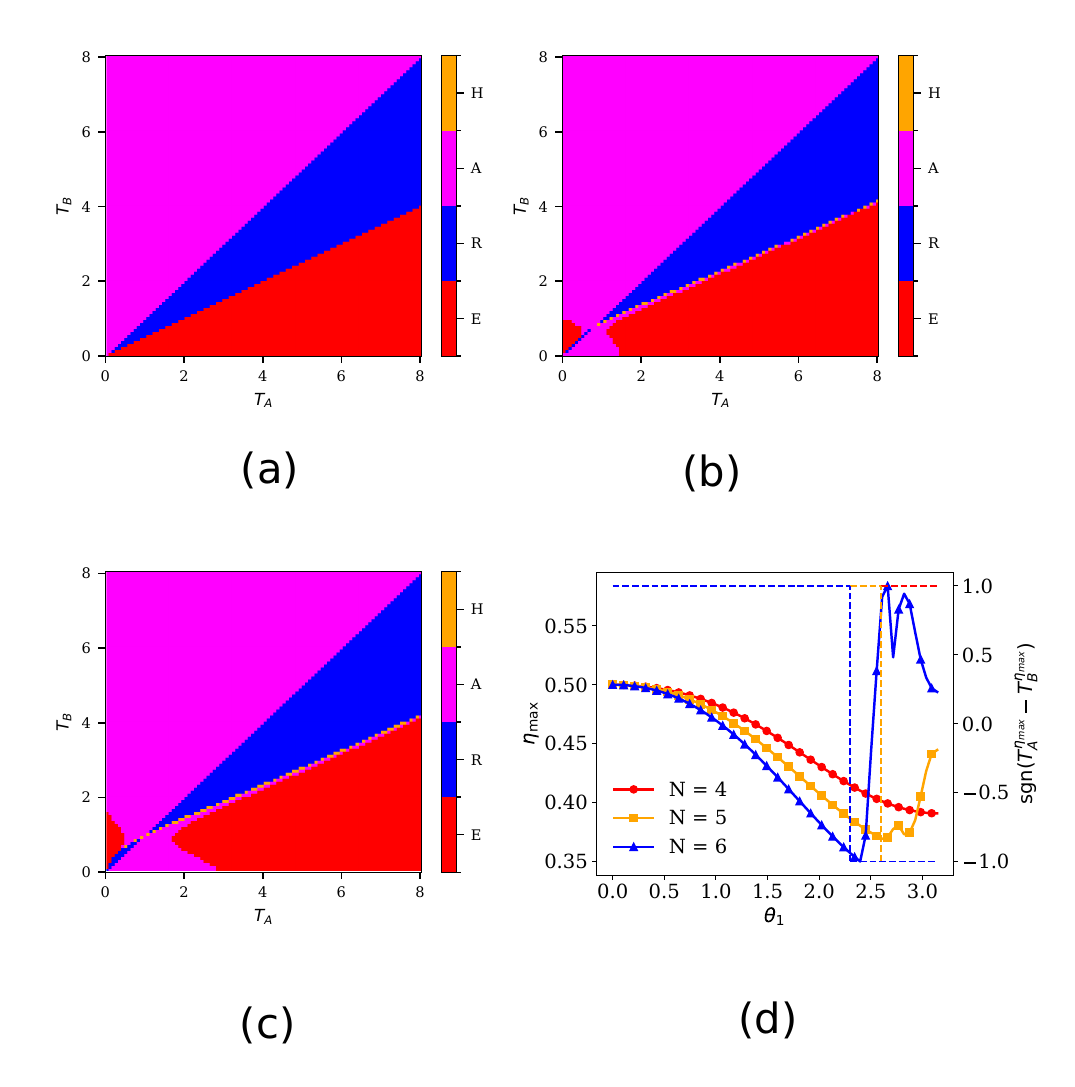}
    \caption{ {\bf Performance of the  HAO cycle corresponding to modified modified work $\overline{W}$ and heat $\overline{Q}$} (see Eq.~\eqref{eq_mod_work_heat} and discussion therein).  Operation modes in the non-interacting limit $U=0$ for $\theta_2=0$, and  (a) $\theta_1=0$, (b) $\theta_1=2.2$ and (c) $\theta_1=\pi$. (d) Maximum efficiency $\eta_{max}$ calculated over $0<T_A,T_B<8$ as a function of $\theta_1$ for $\theta_2=0$. The  dashed lines correspond to $\mathrm{sgn}(T_A^{\eta_{max}}-T_B^{\eta_{max}})$, where $T_{A(B)}^{max}$ are the temperatures of the bath for which the maximum efficiency is obtained for a fixed $\theta_1$. {The numerical data are calculated for $L=12, U=0, J_1=2.0$ and $J_2=1.0$.}}
    \label{fig:modes_modified}
\end{figure*}

{For $N\lesssim L/ 2$, the monotonous reduction of the probability of high occupancy sites in the ground state as a function of $\theta$ is straightforward to understand. At such low fillings, the system minimizes energy by avoiding configurations that have particles on nearby sites as well as multiple particles on same site, so as to minimize the energy cost due to the density-dependent phase associated with the hopping term  in Eq.~\eqref{eq:ham} for $\theta\neq 0$. However, as  $N$ approaches $L/2$ from below, such configurations cannot be avoided. In fact, previous results have indicated a phase transition from superfluid phase to a partially-paired phase as $\theta\to\pi$ for $N\approx L/2$ \cite{anyonhubbard15}. The partially-paired phase in particular has been shown to favor configurations with double occupancy per site, which is consistent with our observation of increasing probability of multiple particle occupations per site as $\theta\to\pi$.    Intuitively, the origin of the above behavior can be attributed to the fact that the density-dependent phase can only acquire the values $\approx\pm 1$ as $\theta\to\pi$, which makes higher occupation per site favorable again, unlike for intermediate values of $\theta$. In \cite{supp}, we provide further numerical evidence in support of this argument.} 

\paragraph*{\bf Reconciliation with the second law}\label{sec:recon}
In the previous sections, we have argued that the apparent violation of the second law of thermodynamics in the IA mode can be reconciled if the work associated with the buildup of anyon energy is accounted for. To show this explicitly, we modify the definition of work in Eq.~\eqref{eq_W12} and Eq.~\eqref{eq_W21} to incorporate the energy change associated with the anyonization and bosonization processes. Likewise, we subtract the same from the heat exchanges to ensure consistency with the first law. Furthermore, in this case, we assume that the  anyonization/bosonization process is carried out much faster than the time-scale of thermalization with the baths. This ensures that energy change during the anyonization/bosonization process is only due to work performed by explicit change of the parameter $\theta$ in the Hamiltonian. Thus, the redefined work and heat are, $\overline{W}_{12(21)}={W}_{12(21)}-\Delta_{B(A)}$, $\overline{W}=\overline{W}_{12} + \overline{W}_{21}$, $\overline{Q}_{A(B)} = Q_{A(B)} - \Delta_{A(B)}$, where 
\begin{equation}\label{eq_mod_work_heat}
\Delta_{B(A)} = \trace \Big[\rho_{2(1)} \Big( H(\theta_{2(1)}, \lambda_{2(1)}) - H(\theta_{1(2)}, \lambda_{2(1)})\Big)\Big].    \end{equation}
Since the HAO cycle operates like a regular Otto cycle in terms of $\overline{W}$ and $\overline{Q}$, it now also becomes meaningful to define the efficiency $\eta=\overline{W}/\overline{Q}_{A(B)}$ for $T_{A(B)}>T_{B(A)}$, provided the cycle is operating in the engine mode.  Examining the non-interacting situation with $U=0$, $J_1=2.0$ and $J_2=1.0$, it is evident from Figs.~\ref{fig:modes_modified} \tb{(a)-(c)} that the inverse accelerator mode does not emerge even for $\theta_1\neq 0$. However, the effect of changing the statistical properties with $\theta_1$ still manifests itself at low temperatures. In particular, for $\theta_1\geq2.2$, the cycle is able to operate as an accelerator for certain range of small but finite temperatures for $T_A>T_B$. Likewise, for $T_A<T_B$, the cycle switches from accelerator to engine mode $\theta_1\geq2.2$ at small temperatures.  Finally, we also note the emergence of heater mode {(see Table.~\ref{table:summary})} in the vicinity of the temperatures at which the direction of heat flow switches for the two baths. 

It is fascinating to note that, the maximum possible efficiency of the cycle when operating in engine mode is significantly enhanced for $\theta_1\geq 2.2$. In fact, the maximum efficiency is found to be from the engine mode at $T_A<T_B$, which emerges for $\theta_1\geq 2.2$ (bottom left in Figs.~\ref{fig:modes_modified}\tb{(b)} and ~\ref{fig:modes_modified}\tb{(c)}. This is illustrated in ~\ref{fig:modes_modified}\tb{(d)}, where we plot the maximum efficiency across all temperatures as a function of $\theta_1$. In the same figure, we also plot $\mathrm{sgn}(T_A^{max}-T_B^{max})$ (dashed lines), where $T_{A(B)}^{max}$ are the temperatures of the bath for which the maximum efficiency is obtained for a fixed $\theta_1$. For $N=6$, we can clearly see that $\eta_{max}$ increases drastically for $\theta_1>2.5$. This coincides with $T_A^{max}$ becoming smaller than $T_B^{max}$, signifying that the higher efficiency is extracted from the newly emerged engine regime at small temperature. 

\paragraph*{\bf Experimental realization}
The 1D AHM has already been realized experimentally using ultracold $^{87}\textrm{Rb}$ atoms in an optical lattice \cite{greinerExpt24}, where the tunneling amplitude is engineered to depend on the particle density in accordance with Eq.~\eqref{eq:ham}. In this section, we propose how the HAO thermal machine can be realized using the same experimental setup. The main challenges associated with implementing the HAO cycle on the optical lattice platform of \cite{greinerExpt24} are the following - (i) realization of a thermal bath with which the system interacts during the thermalization strokes, (ii) measurement of the energy change of the system at the end of each stroke, and (iii) requirement of higher particle density ($N>2$) to observe the superior performance of the cycle for $0<\theta<\pi$.

To overcome the first challenge, we propose to introduce a second set of atoms in the lattice, which acts as a bath for the atoms already present in the system. Specifically, the additional atoms that form the bath will be prepared in the superfluid state ($\theta_{\rm bath}\to 0$) to facilitate a controlled tunability of the effective temperature of the bath. Similar realization of controlled bath environments has been proposed and experimentally implemented in a number of other works \cite{Seetharam_2015,Rubio_Abadal_2019,L_onard_2023,Schnell_2024},  and allows for the controlled engineering of dissipation. In our setup, the bath will be located along the length $L$ of the system, {forming a contiguous region of the lattice adjacent to the system sites}, where it is prepared in the superfluid state using the Floquet scheme described in \cite{Cardarelli_2016}. The size of the bath will be much larger than the system such that the system can couple to continuum modes. Independent control of the statistical parameter $\theta$ will be necessary for the bath and the system, as $\theta_{\rm bath}\to 0$ while $\theta_{\rm system}$ changes throughout the HAO cycle, and can be realized by implementing the Floquet scheme using a digital micromirror device (DMD) rather than directly modulating the intensity of the lattice. {In this approach, both the system and bath are realized using the same set of modulation frequencies, with the bath corresponding to the case $\theta=0$, implemented via spatially dependent control of the relative phases of the drive.} Using the DMD to modulate the intensity allows for spatial control of the statistical parameter, as the DMD can be programmed to {project modulated light with different phase configurations onto contiguous regions of the lattice, thereby realizing the system and the bath, respectively.} The system will be separated from the bath by an optically projected wall, implemented either by the same or another DMD, where the height of the wall is determined by the intensity of the light. By changing the intensity of the light, and therefore the height of the wall, the system can be coupled or isolated from the bath during the thermalization or work strokes, respectively. {During the thermalization strokes, the height of the wall will be controlled so as to suppress leakage of particles into the bath while allowing for energy exchange between the system and the bath.}

{Floquet heating is an important consideration in interacting systems, as energy absorption can in principle lead to population of higher bands and breakdown of the single-band description. In practice, this is mitigated by choosing modulation frequencies that are off-resonant with both the low-energy scales (tunneling and interactions) and the band gap, resulting in a long-lived prethermal regime \cite{sun_optimal_2020}. Experiments realizing the AHM have demonstrated minimal heating and good agreement with single-band predictions over timescales relevant for state preparation and observation \cite{bakkalihassani2026}. Additional techniques, such as multi-frequency driving to suppress interband coupling, may further extend this regime \cite{viebahn2021}.}

Concerning the second challenge, we shall resort to time-of-flight (TOF) measurements to find the energy $E(\lambda, \theta)$ at the end of each stroke. We illustrate the measurement protocol only for the non-interacting case ($U=0$) here, as the same protocol will also work for weak interactions. In the absence of interactions, we recall that the work strokes consist of adiabatically tuning the parameter $J$ as the system evolves unitarily, while the heat strokes correspond to a dissipative evolution of the system in contact with thermal baths along with a simultaneous tuning of the phase parameter $\theta$. The measurement protocol will therefore be as follows. Let us assume that the system is initially prepared in the state $\rho_3$, i.e., in thermal equilibrium with the bath with temperature $T_B$ and $\theta=0$. We first perform a TOF measurement to determine the energy $E(J_2, 0)$ of the system in this state. The system is confined by optically projected walls along both $x$ and $y$, with the walls along $y$ defining the length $L$, and to perform TOF, the walls along $y$ are quenched to allow the system to expand within a 1D tube. Fluorescence images of the atoms are taken at successively longer times after expansion to track the expansion profile of the system, from which a velocity and therefore the energy may be extracted. Since the TOF measurement destroys the initial state, we reset the system to $\rho_3$ before starting the HAO cycle. The compression work stroke is performed by decoupling the bath and adiabatically tuning the hopping amplitude $J_2\to J_1$ over a time-interval $\tau_W$, during which the system evolves to $\rho_4$. At $t=\tau_W$, the TOF measurement is performed again to find the energy $E(J_1, 0)$. The work $W_{21}$ is therefore obtained as $W_{21}=\langle E(J_1, 0) - E(J_2, 0)\rangle$. 

In the next step, we once again prepare the system in state $\rho_3$, evolve unitarily to $\rho_2$, followed by coupling the system to bath with temperature $T_A$. The phase parameter is then tuned to $\theta=\theta_1$ while in contact with the bath, and we ensure that the system thermalizes with the bath for sufficiently long time after tuning of the phase parameter is completed, before performing the TOF measurement to find $E(J_1, \theta_1)$. The heat exchanged is thus calculated as $Q_A=\langle E(J_1, \theta_1) - E(J_2, 0)\rangle - W_{21}$. Similarly, to calculate $W_{12}$ and $Q_B$, we shall reset the system to $\rho_3$ in each case, and then perform TOF measurements at the end of the unitary {compression} and thermalization B strokes, respectively, to find $W_{12}= \langle E(J_2, \theta_1) - E(J_2, 0)\rangle - W_{21} - Q_A$ and $Q_{B}= \langle E(J_2, 0)' - E(J_2, 0)\rangle - W_{21} - Q_A - W_{12}$, where $E(J_2, 0)'$ is the measured energy of the system after completing the cycle. We note that $E(J_2, 0)' \neq E(J_2, 0)$ if the thermalization is not perfect in the heat exchange strokes, which will indicate an incomplete cycle.   

Finally, regarding the third challenge, achieving a higher particle density ($N > 2$) is possible as long as the filling $ N/L \lesssim 0.36$ \cite{Str_ter_2016}; otherwise, the Floquet implementation does not faithfully realize the AHM.
However, we note that the underlying Floquet-engineering framework is expected to generalize to larger particle numbers through the inclusion of additional frequency components in the drive, consistent with related schemes such as that in Ref.~\cite{greinerExpt24}. In particular, higher driving frequencies can be used to resonantly address the interaction-energy scales associated with higher-density configurations arising from larger particle occupations. A systematic exploration of such multi-chromatic driving protocols for higher particle densities is left for future work.
Furthermore, the key features of the HAO cycle, including finite work extraction at low temperatures and its dependence on the statistical parameter do not require a high filling fraction and can be observed even for $N\ll L/2$ . Finally, we note that other implementations that do not involve shaking may provide a promising route toward realizing such systems at higher filling \cite{Fr_lian_2022}.

\section*{Discussion}\label{sec:concl}
In summary, we have proposed a four-stroke hybrid anyon Otto cycle based on the 1D anyon Hubbard model which, like its Pauli counterpart, relies on exclusion statistics of anyons to derive work at low temperatures. In the absence of an explicit interaction among the anyons, we see a monotonous increase in the low temperature work output as the statistical parameter is increased from the bosonic limit to the pseudo-fermionic limit. The presence of a finite anyon energy at low temperature leads to the emergence of an inverse accelerator mode of the cycle, which is prohibited by the second law in a regular Otto cycle. 
When finite but weak interactions are introduced, and at half or higher filling, the low-temperature work output no longer peaks at the bosonic or pseudo-fermionic limits. Instead, it reaches a maximum at an intermediate value of the statistical parameter, thus demonstrating that in the interacting regime, anyonic statistics can be harnessed to achieve greater work extraction than is possible with either bosonic or fermionic statistics alone.
We reiterate that the inverse accelerator mode does not violate the second law in the HAO cycle; they arise from treating the anyonization stroke - which can be interpreted as a work stroke - as a “heat” source, as clarified in the final discussion. By properly incorporating the anyonic contribution and redefining the work, we show that at low temperatures the accelerator mode is replaced by an engine mode. This engine mode exhibits maximum efficiency in the anyonic limit at high $\theta$, highlighting the crucial role of anyonic statistics in enhancing cycle performance. Finally, we outline an experimental protocol to verify our results using a similar experimental platform that was used to realize the AHM. \\

\section*{Methods}
\subsection*{Numerical simulations}
All the numerical simulations were performed in the bosonic representation of the AHM Hamiltonian (Eq.~\eqref{eq:ham}) using the QuSpin package \cite{quspin1, quspin2}. The density dependent phase of the hopping terms cannot be directly implemented using the standard Quspin routines for constructing Hamiltonians. Hence, we explicitly encoded the hopping terms as sparse matrices and added the same to the onsite interaction terms constructed directly using QuSpin routines \cite{code_ahm_hamiltonian}. The full Hamiltonian was then diagonalized within Quspin for generation of required data.

\section*{Data Availability}
All data generated or analysed during this study are included in this article. 

\section*{Code Availability}
The code for constructing the AHM Hamiltonian can be accessed via the link 
https://doi.org/10.5281/zenodo.17873350. 

\section*{Acknowledgements}
ICFO-QOT group acknowledges support from: European Research Council AdG NOQIA; MCIN/AEI (PGC2018-0910.13039/501100011033, CEX2019-000910-S/10.13039/501100011033, Plan National FIDEUA PID2019-106901GB-I00, Plan National STAMEENA PID2022-139099NB, I00, project funded by MCIN/AEI/10.13039/501100011033 and by the “European Union NextGenerationEU/PRTR"; (PRTR-C17.I1), FPI); QUANTERA DYNAMITE PCI2022-132919, QuantERA II Programme co-funded by European Union’s Horizon 2020 program under Grant Agreement No 101017733; Ministry for Digital Transformation and of Civil Service of the Spanish Government through the QUANTUM ENIA project call - Quantum Spain project, and by the European Union through the Recovery, Transformation and Resilience Plan - NextGenerationEU within the framework of the Digital Spain 2026 Agenda; Fundació Cellex; Fundació Mir-Puig; Generalitat de Catalunya (European Social Fund FEDER and CERCA program; Barcelona Supercomputing Center MareNostrum (FI-2023-3-0024); Funded by the European Union. Views and opinions expressed are however those of the author(s) only and do not necessarily reflect those of the European Union, European Commission, European Climate, Infrastructure and Environment Executive Agency (CINEA), or any other granting authority. Neither the European Union nor any granting authority can be held responsible for them (HORIZON-CL4-2022-QUANTUM-02-SGA PASQuanS2.1, 101113690, EU Horizon 2020 FET-OPEN OPTOlogic, Grant No 899794, QU-ATTO, 101168628), EU Horizon Europe Program (This project has received funding from the European Union’s Horizon Europe research and innovation program under grant agreement No 101080086 NeQSTGrant Agreement 101080086 — NeQST); ICFO Internal “QuantumGaudi” project.
R.W.C acknowledges support from the Polish National Science Centre (NCN) under the Maestro Grant No. DEC- 2019/34/A/ST2/00081. 
U.B. is also grateful for the financial support of the IBM Quantum Researcher Program.
T.G. acknowledges the financial support received from the IKUR Strategy under the collaboration agreement between the Ikerbasque Foundation and DIPC on behalf of the Department of Education of the Basque Government, as well as funding by the Department of Education of the Basque Government through the project PIBA\_2023\_1\_0021 (TENINT), and by the Agencia Estatal de Investigación (AEI) through Proyectos de Generación de Conocimiento PID2022-142308NA-I00 (EXQUSMI). A.P. and M.L.B. acknowledge for support from the project PID2023-152724NA-I00, with funding from MCIU/AEI/10.13039/501100011033 and FSE+, the Severo Ochoa Grant CEX2023-001292-S, Generalitat Valenciana grant CIPROM/2022/66, the Ministry of Economic Affairs and Digital Transformation of the Spanish Government through
the QUANTUM ENIA project call - QUANTUM SPAIN project, and by the European Union through the Recovery, Transformation and Resilience Plan - NextGenerationEU within the framework of the Digital Spain 2026 Agenda, and by the CSIC Interdisciplinary Thematic Platform (PTI+) on Quantum Technologies (PTI-QTEP+). Also funding from Horizon Europe EU projects MSCA-SE CaLIGOLA, Project ID: 101086123, and MSCA-DN CaLiForNIA, Project ID: 101119552.   These authors gratefully acknowledge the computer resources at Artemisa, funded by the European Union ERDF and Comunitat Valenciana as well as the technical support provided by the Instituto de Fisica Corpuscular, IFIC (CSIC-UV). 
 M.A G-M acknowledges support  from the Ministry for Digital Transformation and of Civil Service of the Spanish Government through the QUANTUM ENIA project call—Quantum Spain project, and by the European Union through the Recovery, Transformation and Resilience Plan—NextGenerationEU within the framework of the Digital Spain 2026 Agenda: also from Projects of MCIN with funding from European Union NextGenerationEU (PRTR-C17.I1) and by Generalitat Valenciana, with reference 20220883 (PerovsQuTe) and COMCUANTICA/007 (QuanTwin).

\section*{Authors contributions}
SB and UB conceived the project. MB led the numerical calculations, with additional contributions from SB. SB and UB performed the main analysis and interpretation of the results, with further contributions from ML, TG, and RC. AP and MAG-M also contributed to a part of the analysis and interpretation.  JK developed the experimental proposal. All authors contributed to preparation of the manuscript. All authors have read and approved the manuscript.

\section*{Competing Interests}
The authors declare no competing financial or non-financial interests.
\bibliography{ref}

@Article{souravReview21,
author={Bhattacharjee, Sourav
and Dutta, Amit},
title={Quantum thermal machines and batteries},
journal={The European Physical Journal B},
year={2021},
month={Dec},
day={08},
volume={94},
number={12},
pages={239},
issn={1434-6036},
doi={10.1140/epjb/s10051-021-00235-3},
url={https://doi.org/10.1140/epjb/s10051-021-00235-3}
}

@misc{bakkalihassani2026,
      title={Revealing Pseudo-Fermionization and Chiral Binding of One-Dimensional Anyons using Adiabatic State Preparation}, 
      author={Brice Bakkali-Hassani and Joyce Kwan and Perrin Segura and Yanfei Li and Isaac Tesfaye and Gerard Valentí-Rojas and André Eckardt and Markus Greiner},
      year={2026},
      eprint={2602.20421},
      archivePrefix={arXiv},
      primaryClass={cond-mat.quant-gas},
      url={https://arxiv.org/abs/2602.20421}, 
}

@article{sun_optimal_2020,
  title = {Optimal Frequency Window for {{Floquet}} Engineering in Optical Lattices},
  author = {Sun, Gaoyong and Eckardt, Andr{\'e}},
  year = 2020,
  month = mar,
  journal = {Phys. Rev. Res.},
  volume = {2},
  number = {1},
  pages = {013241},
  publisher = {American Physical Society},
  doi = {10.1103/PhysRevResearch.2.013241},
  urldate = {2026-03-10}
}

@article{viebahn2021,
  title = {Suppressing Dissipation in a Floquet-Hubbard System},
  author = {Viebahn, Konrad and Minguzzi, Joaqu\'{\i}n and Sandholzer, Kilian and Walter, Anne-Sophie and Sajnani, Manish and G\"org, Frederik and Esslinger, Tilman},
  journal = {Phys. Rev. X},
  volume = {11},
  issue = {1},
  pages = {011057},
  numpages = {15},
  year = {2021},
  month = {Mar},
  publisher = {American Physical Society},
  doi = {10.1103/PhysRevX.11.011057},
  url = {https://link.aps.org/doi/10.1103/PhysRevX.11.011057}
}

@article{greinerExpt24,
author = {Joyce Kwan  and Perrin Segura  and Yanfei Li  and Sooshin Kim  and Alexey V. Gorshkov  and André Eckardt  and Brice Bakkali-Hassani  and Markus Greiner },
title = {Realization of one-dimensional anyons with arbitrary statistical phase},
journal = {Science},
volume = {386},
number = {6725},
pages = {1055-1060},
year = {2024},
doi = {10.1126/science.adi3252},
URL = {https://www.science.org/doi/abs/10.1126/science.adi3252},
eprint = {https://www.science.org/doi/pdf/10.1126/science.adi3252}}

@Article{Koch2023,
author={Koch, Jennifer
and Menon, Keerthy
and Cuestas, Eloisa
and Barbosa, Sian
and Lutz, Eric
and Fogarty, Thom{\'a}s
and Busch, Thomas
and Widera, Artur},
title={A quantum engine in the BEC--BCS crossover},
journal={Nature},
year={2023},
month={Sep},
day={01},
volume={621},
number={7980},
pages={723-727},
issn={1476-4687},
doi={10.1038/s41586-023-06469-8},
url={https://doi.org/10.1038/s41586-023-06469-8}
}

@Article{Dhar2025,
author={Dhar, Sudipta
and Wang, Botao
and Horvath, Milena
and Vashisht, Amit
and Zeng, Yi
and Zvonarev, Mikhail B.
and Goldman, Nathan
and Guo, Yanliang
and Landini, Manuele
and N{\"a}gerl, Hanns-Christoph},
title={Observing anyonization of bosons in a quantum gas},
journal={Nature},
year={2025},
month={Jun},
day={01},
volume={642},
number={8066},
pages={53-57},
issn={1476-4687},
doi={10.1038/s41586-025-09016-9},
url={https://doi.org/10.1038/s41586-025-09016-9}
}

@misc{manianyoncarnot25,
      title={The Anyonic Quantum Carnot Engine}, 
      author={H S Mani and N Ramadas and V V Sreedhar},
      year={2025},
      eprint={2504.20596},
      archivePrefix={arXiv},
      primaryClass={quant-ph},
      url={https://arxiv.org/abs/2504.20596}, 
}

@misc{manianyonotto24,
      title={Quantum Thermodynamics of Small Systems: The Anyonic Otto Engine}, 
      author={H S Mani and Ramadas N and V V Sreedhar},
      year={2024},
      eprint={2401.07177},
      archivePrefix={arXiv},
      primaryClass={quant-ph},
      url={https://arxiv.org/abs/2401.07177}, 
}

@article{anyonhubbard15,
  title = {Anyon Hubbard Model in One-Dimensional Optical Lattices},
  author = {Greschner, Sebastian and Santos, Luis},
  journal = {Phys. Rev. Lett.},
  volume = {115},
  issue = {5},
  pages = {053002},
  numpages = {5},
  year = {2015},
  month = {Jul},
  publisher = {American Physical Society},
  doi = {10.1103/PhysRevLett.115.053002},
  url = {https://link.aps.org/doi/10.1103/PhysRevLett.115.053002}
}

@article{Rubio_Abadal_2019,
   title={Many-Body Delocalization in the Presence of a Quantum Bath},
   volume={9},
   ISSN={2160-3308},
   url={http://dx.doi.org/10.1103/PhysRevX.9.041014},
   DOI={10.1103/physrevx.9.041014},
   number={4},
   journal={Physical Review X},
   publisher={American Physical Society (APS)},
   author={Rubio-Abadal, Antonio and Choi, Jae-yoon and Zeiher, Johannes and Hollerith, Simon and Rui, Jun and Bloch, Immanuel and Gross, Christian},
   year={2019},
   month=oct 
}

@article{L_onard_2023,
   title={Probing the onset of quantum avalanches in a many-body localized system},
   volume={19},
   ISSN={1745-2481},
   url={http://dx.doi.org/10.1038/s41567-022-01887-3},
   DOI={10.1038/s41567-022-01887-3},
   number={4},
   journal={Nature Physics},
   publisher={Springer Science and Business Media LLC},
   author={Léonard, Julian and Kim, Sooshin and Rispoli, Matthew and Lukin, Alexander and Schittko, Robert and Kwan, Joyce and Demler, Eugene and Sels, Dries and Greiner, Markus},
   year={2023},
   month=jan, pages={481–485} 
}

@article{Cardarelli_2016,
   title={Engineering interactions and anyon statistics by multicolor lattice-depth modulations},
   volume={94},
   ISSN={2469-9934},
   url={http://dx.doi.org/10.1103/PhysRevA.94.023615},
   DOI={10.1103/physreva.94.023615},
   number={2},
   journal={Physical Review A},
   publisher={American Physical Society (APS)},
   author={Cardarelli, Lorenzo and Greschner, Sebastian and Santos, Luis},
   year={2016},
   month=aug 
}

@article{Str_ter_2016,
   title={Floquet Realization and Signatures of One-Dimensional Anyons in an Optical Lattice},
   volume={117},
   ISSN={1079-7114},
   url={http://dx.doi.org/10.1103/PhysRevLett.117.205303},
   DOI={10.1103/physrevlett.117.205303},
   number={20},
   journal={Physical Review Letters},
   publisher={American Physical Society (APS)},
   author={Sträter, Christoph and Srivastava, Shashi C.L. and Eckardt, André},
   year={2016},
   month=nov 
}

@article{Fr_lian_2022,
   title={Realizing a 1D topological gauge theory in an optically dressed BEC},
   volume={608},
   ISSN={1476-4687},
   url={http://dx.doi.org/10.1038/s41586-022-04943-3},
   DOI={10.1038/s41586-022-04943-3},
   number={7922},
   journal={Nature},
   publisher={Springer Science and Business Media LLC},
   author={Frölian, Anika and Chisholm, Craig S. and Neri, Elettra and Cabrera, Cesar R. and Ramos, Ramón and Celi, Alessio and Tarruell, Leticia},
   year={2022},
   month=aug, pages={293–297} 
}

@Article{keilmann11,
author={Keilmann, Tassilo
and Lanzmich, Simon
and McCulloch, Ian
and Roncaglia, Marco},
title={Statistically induced phase transitions and anyons in 1D optical lattices},
journal={Nature Communications},
year={2011},
month={Jun},
day={21},
volume={2},
number={1},
pages={361},
issn={2041-1723},
doi={10.1038/ncomms1353},
url={https://doi.org/10.1038/ncomms1353}
}

@article{Seetharam_2015,
   title={Controlled Population of Floquet-Bloch States via Coupling to Bose and Fermi Baths},
   volume={5},
   ISSN={2160-3308},
   url={http://dx.doi.org/10.1103/PhysRevX.5.041050},
   DOI={10.1103/physrevx.5.041050},
   number={4},
   journal={Physical Review X},
   publisher={American Physical Society (APS)},
   author={Seetharam, Karthik I. and Bardyn, Charles-Edouard and Lindner, Netanel H. and Rudner, Mark S. and Refael, Gil},
   year={2015},
   month=dec 
}

@article{Schnell_2024,
   title={Dissipative preparation of a Floquet topological insulator in an optical lattice via bath engineering},
   volume={17},
   ISSN={2542-4653},
   url={http://dx.doi.org/10.21468/SciPostPhys.17.2.052},
   DOI={10.21468/scipostphys.17.2.052},
   number={2},
   journal={SciPost Physics},
   publisher={Stichting SciPost},
   author={Schnell, Alexander and Weitenberg, Christof and Eckardt, André},
   year={2024},
   month=aug 
}

@misc{uusnäkki25experiment,
      title={Experimental realization of a quantum heat engine based on dissipation-engineered superconducting circuits}, 
      author={Tuomas Uusnäkki and Timm Mörstedt and Wallace Teixeira and Miika Rasola and Mikko Möttönen},
      year={2025},
      eprint={2502.20143},
      archivePrefix={arXiv},
      primaryClass={quant-ph},
      url={https://arxiv.org/abs/2502.20143}, 
}

@article{Aamir2025,
author={Aamir, Mohammed Ali
and Jamet Suria, Paul
and Mar{\'i}n Guzm{\'a}n, Jos{\'e} Antonio
and Castillo-Moreno, Claudia
and Epstein, Jeffrey M.
and Yunger Halpern, Nicole
and Gasparinetti, Simone},
title={Thermally driven quantum refrigerator autonomously resets a superconducting qubit},
journal={Nature Physics},
year={2025},
month={Feb},
day={01},
volume={21},
number={2},
pages={318-323},
issn={1745-2481},
doi={10.1038/s41567-024-02708-5},
url={https://doi.org/10.1038/s41567-024-02708-5}
}

@article{gluza21,
  title = {Quantum Field Thermal Machines},
  author = {Gluza, Marek and Sabino, Jo\~ao and Ng, Nelly H.Y. and Vitagliano, Giuseppe and Pezzutto, Marco and Omar, Yasser and Mazets, Igor and Huber, Marcus and Schmiedmayer, J\"org and Eisert, Jens},
  journal = {PRX Quantum},
  volume = {2},
  issue = {3},
  pages = {030310},
  numpages = {48},
  year = {2021},
  month = {Jul},
  publisher = {American Physical Society},
  doi = {10.1103/PRXQuantum.2.030310},
  url = {https://link.aps.org/doi/10.1103/PRXQuantum.2.030310}
}

@article{lutz14,
  title = {Nanoscale Heat Engine Beyond the Carnot Limit},
  author = {Ro\ss{}nagel, J. and Abah, O. and Schmidt-Kaler, F. and Singer, K. and Lutz, E.},
  journal = {Phys. Rev. Lett.},
  volume = {112},
  issue = {3},
  pages = {030602},
  numpages = {5},
  year = {2014},
  month = {Jan},
  publisher = {American Physical Society},
  doi = {10.1103/PhysRevLett.112.030602},
  url = {https://link.aps.org/doi/10.1103/PhysRevLett.112.030602}
}

@article{arezzo24,
  title = {Many-body quantum heat engines based on free fermion systems},
  author = {Arezzo, Vincenzo Roberto and Rossini, Davide and Piccitto, Giulia},
  journal = {Phys. Rev. B},
  volume = {109},
  issue = {22},
  pages = {224309},
  numpages = {15},
  year = {2024},
  month = {Jun},
  publisher = {American Physical Society},
  doi = {10.1103/PhysRevB.109.224309},
  url = {https://link.aps.org/doi/10.1103/PhysRevB.109.224309}
}

@article{hartmann20,
  title = {Many-body quantum heat engines with shortcuts to adiabaticity},
  author = {Hartmann, Andreas and Mukherjee, Victor and Niedenzu, Wolfgang and Lechner, Wolfgang},
  journal = {Phys. Rev. Res.},
  volume = {2},
  issue = {2},
  pages = {023145},
  numpages = {13},
  year = {2020},
  month = {May},
  publisher = {American Physical Society},
  doi = {10.1103/PhysRevResearch.2.023145},
  url = {https://link.aps.org/doi/10.1103/PhysRevResearch.2.023145}
}

@Article{Niedenzu18,
author={Niedenzu, Wolfgang
and Mukherjee, Victor
and Ghosh, Arnab
and Kofman, Abraham G.
and Kurizki, Gershon},
title={Quantum engine efficiency bound beyond the second law of thermodynamics},
journal={Nature Communications},
year={2018},
month={Jan},
day={11},
volume={9},
number={1},
pages={165},
issn={2041-1723},
doi={10.1038/s41467-017-01991-6},
url={https://doi.org/10.1038/s41467-017-01991-6}
}

@article{kosloff14,
   author = "Kosloff, Ronnie and Levy, Amikam",
   title = "Quantum Heat Engines and Refrigerators: Continuous Devices", 
   journal= "Annual Review of Physical Chemistry",
   year = "2014",
   volume = "65",
   number = "Volume 65, 2014",
   pages = "365-393",
   doi = "https://doi.org/10.1146/annurev-physchem-040513-103724",
   url = "https://www.annualreviews.org/content/journals/10.1146/annurev-physchem-040513-103724",
   publisher = "Annual Reviews",
   issn = "1545-1593",
   type = "Journal Article",
  }

@Article{mukherjee24,
author={Mukherjee, Victor
and Divakaran, Uma},
title={The promises and challenges of many-body quantum technologies: A focus on quantum engines},
journal={Nature Communications},
year={2024},
month={Apr},
day={12},
volume={15},
number={1},
pages={3170},
issn={2041-1723},
doi={10.1038/s41467-024-47638-1},
url={https://doi.org/10.1038/s41467-024-47638-1}
}

@article{revathy20,
  title = {Universal finite-time thermodynamics of many-body quantum machines from Kibble-Zurek scaling},
  author = {B. S, Revathy and Mukherjee, Victor and Divakaran, Uma and del Campo, Adolfo},
  journal = {Phys. Rev. Res.},
  volume = {2},
  issue = {4},
  pages = {043247},
  numpages = {14},
  year = {2020},
  month = {Nov},
  publisher = {American Physical Society},
  doi = {10.1103/PhysRevResearch.2.043247},
  url = {https://link.aps.org/doi/10.1103/PhysRevResearch.2.043247}
}

@article{watanabe20,
  title = {Quantum Statistical Enhancement of the Collective Performance of Multiple Bosonic Engines},
  author = {Watanabe, Gentaro and Venkatesh, B. Prasanna and Talkner, Peter and Hwang, Myung-Joong and del Campo, Adolfo},
  journal = {Phys. Rev. Lett.},
  volume = {124},
  issue = {21},
  pages = {210603},
  numpages = {6},
  year = {2020},
  month = {May},
  publisher = {American Physical Society},
  doi = {10.1103/PhysRevLett.124.210603},
  url = {https://link.aps.org/doi/10.1103/PhysRevLett.124.210603}
}

@article{myers20,
  title = {Bosons outperform fermions: The thermodynamic advantage of symmetry},
  author = {Myers, Nathan M. and Deffner, Sebastian},
  journal = {Phys. Rev. E},
  volume = {101},
  issue = {1},
  pages = {012110},
  numpages = {11},
  year = {2020},
  month = {Jan},
  publisher = {American Physical Society},
  doi = {10.1103/PhysRevE.101.012110},
  url = {https://link.aps.org/doi/10.1103/PhysRevE.101.012110}
}

@article{von19,
  title = {Spin Heat Engine Coupled to a Harmonic-Oscillator Flywheel},
  author = {von Lindenfels, D. and Gr\"ab, O. and Schmiegelow, C. T. and Kaushal, V. and Schulz, J. and Mitchison, Mark T. and Goold, John and Schmidt-Kaler, F. and Poschinger, U. G.},
  journal = {Phys. Rev. Lett.},
  volume = {123},
  issue = {8},
  pages = {080602},
  numpages = {6},
  year = {2019},
  month = {Aug},
  publisher = {American Physical Society},
  doi = {10.1103/PhysRevLett.123.080602},
  url = {https://link.aps.org/doi/10.1103/PhysRevLett.123.080602}
}

@misc{ray23,
      title={Kerr-type nonlinear baths enhance cooling in quantum refrigerators}, 
      author={Tanaya Ray and Sayan Mondal and Aparajita Bhattacharyya and Ahana Ghoshal and Debraj Rakshit and Ujjwal Sen},
      year={2023},
      eprint={2311.10499},
      archivePrefix={arXiv},
      primaryClass={quant-ph},
      url={https://arxiv.org/abs/2311.10499}, 
}

@article{sen25,
  title = {Transient effects in quantum refrigerators with finite environments},
  author = {Bhattacharyya, Aparajita and Ghoshal, Ahana and Sen, Ujjwal},
  journal = {Phys. Rev. A},
  volume = {111},
  issue = {1},
  pages = {012209},
  numpages = {18},
  year = {2025},
  month = {Jan},
  publisher = {American Physical Society},
  doi = {10.1103/PhysRevA.111.012209},
  url = {https://link.aps.org/doi/10.1103/PhysRevA.111.012209}
}

@article{scully03,
  title = {Extracting work from a single heat bath via vanishing quantum coherence},
  author = {Scully, Marlan O and Zubairy, M Suhail and Agarwal, Girish S and Walther, Herbert},
  journal  = {Science},
  volume   =  {299},
  number   =  {5608},
  pages    = {862--864},
  month    =  {jan},
  year     =  {2003},
  address  = {United States}
}

@article{klaers17,
  title = {Squeezed Thermal Reservoirs as a Resource for a Nanomechanical Engine beyond the Carnot Limit},
  author = {Klaers, Jan and Faelt, Stefan and Imamoglu, Atac and Togan, Emre},
  journal = {Phys. Rev. X},
  volume = {7},
  issue = {3},
  pages = {031044},
  numpages = {6},
  year = {2017},
  month = {Sep},
  publisher = {American Physical Society},
  doi = {10.1103/PhysRevX.7.031044},
  url = {https://link.aps.org/doi/10.1103/PhysRevX.7.031044}
}

@article{manzano16,
  title = {Entropy production and thermodynamic power of the squeezed thermal reservoir},
  author = {Manzano, Gonzalo and Galve, Fernando and Zambrini, Roberta and Parrondo, Juan M. R.},
  journal = {Phys. Rev. E},
  volume = {93},
  issue = {5},
  pages = {052120},
  numpages = {10},
  year = {2016},
  month = {May},
  publisher = {American Physical Society},
  doi = {10.1103/PhysRevE.93.052120},
  url = {https://link.aps.org/doi/10.1103/PhysRevE.93.052120}
}

@article{camati19,
  title = {Coherence effects in the performance of the quantum Otto heat engine},
  author = {Camati, Patrice A. and Santos, Jonas F. G. and Serra, Roberto M.},
  journal = {Phys. Rev. A},
  volume = {99},
  issue = {6},
  pages = {062103},
  numpages = {15},
  year = {2019},
  month = {Jun},
  publisher = {American Physical Society},
  doi = {10.1103/PhysRevA.99.062103},
  url = {https://link.aps.org/doi/10.1103/PhysRevA.99.062103}
}

@Article{hardal15,
author={Hardal, Ali {\"U}. C.
and M{\"u}stecapl{\i}o{\u{g}}lu, {\"O}zg{\"u}r E.},
title={Superradiant Quantum Heat Engine},
journal={Scientific Reports},
year={2015},
month={Aug},
day={11},
volume={5},
number={1},
pages={12953},
issn={2045-2322},
doi={10.1038/srep12953},
url={https://doi.org/10.1038/srep12953}
}

@article{myers22,
  title        = {Boosting engine performance with Bose--Einstein condensation},
  author       = {Nathan M. Myers and Francisco J. Peña and Oscar Negrete and Patricio Vargas and Gabriele De Chiara and Sebastian Deffner},
  journal      = {New Journal of Physics},
  volume       = {24},
  number       = {2},
  pages        = {025001},
  year         = {2022},
  publisher    = {IOP Publishing},
  doi          = {10.1088/1367-2630/ac47cc},
  note         = {Focus on Microscopic Engines and Refrigerators: Theory and Experiments from Classical to Quantum}
}

@article{solfanelli23,
  title        = {Quantum heat engine with long-range advantages},
  author       = {Andrea Solfanelli and Guido Giachetti and Michele Campisi and Stefano Ruffo and Nicol\`o Defenu},
  journal      = {New Journal of Physics},
  volume       = {25},
  number       = {3},
  pages        = {033030},
  year         = {2023},
  publisher    = {IOP Publishing},
  doi          = {10.1088/1367-2630/acc04e}
}

@article{piccitto22,
  title        = {The Ising critical quantum Otto engine},
  author       = {Giulia Piccitto and Michele Campisi and Davide Rossini},
  journal      = {New Journal of Physics},
  volume       = {24},
  number       = {10},
  pages        = {103023},
  year         = {2022},
  publisher    = {IOP Publishing},
  doi          = {10.1088/1367-2630/ac963b}
}

@Article{hazzard25,
author={Wang, Zhiyuan
and Hazzard, Kaden R. A.},
title={Particle exchange statistics beyond fermions and bosons},
journal={Nature},
year={2025},
month={Jan},
day={01},
volume={637},
number={8045},
pages={314-318},
issn={1476-4687},
doi={10.1038/s41586-024-08262-7},
url={https://doi.org/10.1038/s41586-024-08262-7}
}

@article{anyonRev08,
  title = {Non-Abelian anyons and topological quantum computation},
  author = {Nayak, Chetan and Simon, Steven H. and Stern, Ady and Freedman, Michael and Das Sarma, Sankar},
  journal = {Rev. Mod. Phys.},
  volume = {80},
  issue = {3},
  pages = {1083--1159},
  numpages = {0},
  year = {2008},
  month = {Sep},
  publisher = {American Physical Society},
  doi = {10.1103/RevModPhys.80.1083},
  url = {https://link.aps.org/doi/10.1103/RevModPhys.80.1083}
}

@article{sternRev08,
title = {Anyons and the quantum Hall effect—A pedagogical review},
journal = {Annals of Physics},
volume = {323},
number = {1},
pages = {204-249},
year = {2008},
note = {January Special Issue 2008},
issn = {0003-4916},
doi = {https://doi.org/10.1016/j.aop.2007.10.008},
url = {https://www.sciencedirect.com/science/article/pii/S0003491607001674},
author = {Ady Stern}
}

@misc{Dunlop25,
      title={Thermodynamics of Hamiltonian anyons with applications to quantum heat engines}, 
      author={Dunlop,  Joe and Tejero,  Alvaro and Skotiniotis,  Michalis and Manzano,  Daniel},
      year={2025},
      eprint={2502.19019},
      archivePrefix={arXiv},
      primaryClass={quant-ph},
      url={https://arxiv.org/abs/2502.19019}, 
}

@article{Myers2021,
  title = {Thermodynamics of Statistical Anyons},
  volume = {2},
  ISSN = {2691-3399},
  url = {http://dx.doi.org/10.1103/PRXQuantum.2.040312},
  DOI = {10.1103/prxquantum.2.040312},
  number = {4},
  journal = {PRX Quantum},
  publisher = {American Physical Society (APS)},
  author = {Myers,  Nathan M. and Deffner,  Sebastian},
  year = {2021},
  month = oct 
}

@misc{Wang25,
Author = {Botao Wang and Amit Vashisht and Yanliang Guo and Sudipta Dhar and Manuele Landini and Hanns-Christoph Nägerl and Nathan Goldman},
Title = {Anyonization of bosons in one dimension: an effective swap model},
Year = {2025},
Eprint = {arXiv:2504.21208},
}

@article{Bonkhoff2025,
  title = {Anyonic Phase Transitions in the 1D Extended Hubbard Model with Fractional Statistics},
  volume = {135},
  ISSN = {1079-7114},
  url = {http://dx.doi.org/10.1103/7n1c-vq2p},
  DOI = {10.1103/7n1c-vq2p},
  number = {3},
  journal = {Physical Review Letters},
  publisher = {American Physical Society (APS)},
  author = {Bonkhoff,  Martin and J\"{a}gering,  Kevin and Hu,  Shijie and Pelster,  Axel and Eggert,  Sebastian and Schneider,  Imke},
  year = {2025},
  month = jul 
}

@article{Zhang20,
  title = {Statistically related many-body localization in the one-dimensional anyon Hubbard model},
  volume = {102},
  ISSN = {2469-9969},
  url = {http://dx.doi.org/10.1103/PhysRevB.102.054204},
  DOI = {10.1103/physrevb.102.054204},
  number = {5},
  journal = {Physical Review B},
  publisher = {American Physical Society (APS)},
  author = {Zhang,  Guo-Qing and Zhang,  Dan-Wei and Li,  Zhi and Wang,  Z. D. and Zhu,  Shi-Liang},
  year = {2020},
  month = aug 
}

@article{Tang15,
  title = {Ground-state properties of anyons in a one-dimensional lattice},
  volume = {17},
  ISSN = {1367-2630},
  url = {http://dx.doi.org/10.1088/1367-2630/17/12/123016},
  DOI = {10.1088/1367-2630/17/12/123016},
  number = {12},
  journal = {New Journal of Physics},
  publisher = {IOP Publishing},
  author = {Tang,  Guixin and Eggert,  Sebastian and Pelster,  Axel},
  year = {2015},
  month = dec,
  pages = {123016}
}

@article{Kundu99,
  title = {Exact Solution of Double  delta-Function Bose Gas through an Interacting Anyon Gas},
  volume = {83},
  ISSN = {1079-7114},
  url = {http://dx.doi.org/10.1103/PhysRevLett.83.1275},
  DOI = {10.1103/physrevlett.83.1275},
  number = {7},
  journal = {Physical Review Letters},
  publisher = {American Physical Society (APS)},
  author = {Kundu,  Anjan},
  year = {1999},
  month = aug,
  pages = {1275–1278}
}

@article{Zhang17,
  title = {Ground-state properties of the one-dimensional unconstrained pseudo-anyon Hubbard model},
  volume = {95},
  ISSN = {2469-9934},
  url = {http://dx.doi.org/10.1103/PhysRevA.95.053614},
  DOI = {10.1103/physreva.95.053614},
  number = {5},
  journal = {Physical Review A},
  publisher = {American Physical Society (APS)},
  author = {Zhang,  Wanzhou and Greschner,  Sebastian and Fan,  Ernv and Scott,  Tony C. and Zhang,  Yunbo},
  year = {2017},
  month = may 
}

@article{Longhi12,
  title = {Anyons in one-dimensional lattices: a photonic realization},
  volume = {37},
  ISSN = {1539-4794},
  url = {http://dx.doi.org/10.1364/OL.37.002160},
  DOI = {10.1364/ol.37.002160},
  number = {11},
  journal = {Optics Letters},
  publisher = {Optica Publishing Group},
  author = {Longhi,  Stefano and Della Valle,  Giuseppe},
  year = {2012},
  month = jun,
  pages = {2160}
}

@article{Bonkhoff21,
  title = {Bosonic Continuum Theory of One-Dimensional Lattice Anyons},
  volume = {126},
  ISSN = {1079-7114},
  url = {http://dx.doi.org/10.1103/PhysRevLett.126.163201},
  DOI = {10.1103/physrevlett.126.163201},
  number = {16},
  journal = {Physical Review Letters},
  publisher = {American Physical Society (APS)},
  author = {Bonkhoff,  Martin and J\"{a}gering,  Kevin and Eggert,  Sebastian and Pelster,  Axel and Thorwart,  Michael and Posske,  Thore},
  year = {2021},
  month = apr 
}

@article{Theel25,
  title = {Chirally Protected State Manipulation by Tuning One-Dimensional Statistics},
  volume = {135},
  ISSN = {1079-7114},
  url = {http://dx.doi.org/10.1103/kzf6-yx24},
  DOI = {10.1103/kzf6-yx24},
  number = {6},
  journal = {Physical Review Letters},
  publisher = {American Physical Society (APS)},
  author = {Theel,  F. and Bonkhoff,  M. and Schmelcher,  P. and Posske,  T. and Harshman,  N. L.},
  year = {2025},
  month = aug 
}

@article{Bonkhoff23,
  title = {Coherence properties of the repulsive anyon-Hubbard dimer},
  volume = {108},
  ISSN = {2469-9969},
  url = {http://dx.doi.org/10.1103/PhysRevB.108.155134},
  DOI = {10.1103/physrevb.108.155134},
  number = {15},
  journal = {Physical Review B},
  publisher = {American Physical Society (APS)},
  author = {Bonkhoff,  Martin and J\"{a}ger,  Simon B. and Schneider,  Imke and Pelster,  Axel and Eggert,  Sebastian},
  year = {2023},
  month = oct 
}

@article{ArcilaForero16,
  title = {Density matrix renormalization group study of the Anyon-Hubbard model},
  volume = {687},
  ISSN = {1742-6596},
  url = {http://dx.doi.org/10.1088/1742-6596/687/1/012064},
  DOI = {10.1088/1742-6596/687/1/012064},
  journal = {Journal of Physics: Conference Series},
  publisher = {IOP Publishing},
  author = {Arcila-Forero,  J and Franco,  R and Silva-Valencia,  J},
  year = {2016},
  month = feb,
  pages = {012064}
}

@misc{Mostaan25,
      title={Anyon-trions in atomically thin semiconductor heterostructures}, 
      author={Nader Mostaan and Nathan Goldman and Ataç İmamoğlu and Fabian Grusdt},
      year={2025},
      eprint={2507.08933},
      archivePrefix={arXiv},
      primaryClass={cond-mat.mes-hall},
      url={https://arxiv.org/abs/2507.08933}, 
}

@article{Iqbal24,
  title = {Non-Abelian topological order and anyons on a trapped-ion processor},
  volume = {626},
  ISSN = {1476-4687},
  url = {http://dx.doi.org/10.1038/s41586-023-06934-4},
  DOI = {10.1038/s41586-023-06934-4},
  number = {7999},
  journal = {Nature},
  publisher = {Springer Science and Business Media LLC},
  author = {Iqbal,  Mohsin and Tantivasadakarn,  Nathanan and Verresen,  Ruben and Campbell,  Sara L. and Dreiling,  Joan M. and Figgatt,  Caroline and Gaebler,  John P. and Johansen,  Jacob and Mills,  Michael and Moses,  Steven A. and Pino,  Juan M. and Ransford,  Anthony and Rowe,  Mary and Siegfried,  Peter and Stutz,  Russell P. and Foss-Feig,  Michael and Vishwanath,  Ashvin and Dreyer,  Henrik},
  year = {2024},
  month = feb,
  pages = {505–511}
}

@article{Zhang25,
  title = {Fractional-statistics-induced entanglement from Andreev-like tunneling},
  volume = {16},
  ISSN = {2041-1723},
  url = {http://dx.doi.org/10.1038/s41467-025-61869-w},
  DOI = {10.1038/s41467-025-61869-w},
  number = {1},
  journal = {Nature Communications},
  publisher = {Springer Science and Business Media LLC},
  author = {Zhang,  Gu and Glidic,  Pierre and Pierre,  Frédéric and Gornyi,  Igor and Gefen,  Yuval},
  year = {2025},
  month = jul 
}

@misc{Himanshu25,
      title={Aharonov-Bohm Interference in Even-Denominator Fractional Quantum Hall States}, 
      author={Jehyun Kim and Himanshu Dev and Amit Shaer and Ravi Kumar and Alexey Ilin and André Haug and Shelly Iskoz and Kenji Watanabe and Takashi Taniguchi and David F. Mross and Ady Stern and Yuval Ronen},
      year={2024},
      eprint={2412.19886},
      archivePrefix={arXiv},
      primaryClass={cond-mat.mes-hall},
      url={https://arxiv.org/abs/2412.19886}, 
}

@article{Ghosh25,
  title = {Coherent bunching of anyons and dissociation in an interference experiment},
  volume = {642},
  ISSN = {1476-4687},
  url = {http://dx.doi.org/10.1038/s41586-025-09143-3},
  DOI = {10.1038/s41586-025-09143-3},
  number = {8069},
  journal = {Nature},
  publisher = {Springer Science and Business Media LLC},
  author = {Ghosh,  Bikash and Labendik,  Maria and Umansky,  Vladimir and Heiblum,  Moty and Mross,  David F.},
  year = {2025},
  month = jun,
  pages = {922–927}
}

@article{Yutushui25,
  title = {Non-Abelian Phases from the Condensation of Abelian Anyons},
  volume = {135},
  ISSN = {1079-7114},
  url = {http://dx.doi.org/10.1103/3yvl-4hws},
  DOI = {10.1103/3yvl-4hws},
  number = {5},
  journal = {Physical Review Letters},
  publisher = {American Physical Society (APS)},
  author = {Yutushui,  Misha and Hermanns,  Maria and Mross,  David F.},
  year = {2025},
  month = jul 
}

@article{Bartolomei20,
  title = {Fractional statistics in anyon collisions},
  volume = {368},
  ISSN = {1095-9203},
  url = {http://dx.doi.org/10.1126/science.aaz5601},
  DOI = {10.1126/science.aaz5601},
  number = {6487},
  journal = {Science},
  publisher = {American Association for the Advancement of Science (AAAS)},
  author = {Bartolomei,  H. and Kumar,  M. and Bisognin,  R. and Marguerite,  A. and Berroir,  J.-M. and Bocquillon,  E. and Pla\c{c}ais,  B. and Cavanna,  A. and Dong,  Q. and Gennser,  U. and Jin,  Y. and Fève,  G.},
  year = {2020},
  month = apr,
  pages = {173–177}
}

@article{souza22,
  title = {Collective effects on the performance and stability of quantum heat engines},
  author = {Souza, Leonardo da Silva and Manzano, Gonzalo and Fazio, Rosario and Iemini, Fernando},
  journal = {Phys. Rev. E},
  volume = {106},
  issue = {1},
  pages = {014143},
  numpages = {20},
  year = {2022},
  month = {Jul},
  publisher = {American Physical Society},
  doi = {10.1103/PhysRevE.106.014143},
  url = {https://link.aps.org/doi/10.1103/PhysRevE.106.014143}
}

@Article{campisi16critical,
author={Campisi, Michele
and Fazio, Rosario},
title={The power of a critical heat engine},
journal={Nature Communications},
year={2016},
month={Jun},
day={20},
volume={7},
number={1},
pages={11895},
issn={2041-1723},
doi={10.1038/ncomms11895},
url={https://doi.org/10.1038/ncomms11895}
}

@article{fazio16,
doi = {10.1088/1751-8113/49/34/345002},
url = {https://dx.doi.org/10.1088/1751-8113/49/34/345002},
year = {2016},
month = {jul},
publisher = {IOP Publishing},
volume = {49},
number = {34},
pages = {345002},
author = {Campisi, Michele and Fazio, Rosario},
title = {Dissipation, correlation and lags in heat engines},
journal = {Journal of Physics A: Mathematical and Theoretical}
}

@misc{brollo25,
      title={Universal efficiency boost in prethermal quantum heat engines}, 
      author={Alberto Brollo and Adolfo del Campo and Alvise Bastianello},
      year={2025},
      eprint={2504.02044},
      archivePrefix={arXiv},
      primaryClass={quant-ph},
      url={https://arxiv.org/abs/2504.02044}, 
}

@Article{ghosh23,
AUTHOR = {Sur, Saikat and Ghosh, Arnab},
TITLE = {Quantum Advantage of Thermal Machines with Bose and Fermi Gases},
JOURNAL = {Entropy},
VOLUME = {25},
YEAR = {2023},
NUMBER = {2},
ARTICLE-NUMBER = {372},
URL = {https://www.mdpi.com/1099-4300/25/2/372},
PubMedID = {36832738},
ISSN = {1099-4300},
DOI = {10.3390/e25020372}
}

@article{grass20,
  title = {Fractional Angular Momentum and Anyon Statistics of Impurities in Laughlin Liquids},
  author = {Gra\ss{}, Tobias and Juli\'a-D\'{\i}az, Bruno and Baldelli, Niccol\`o and Bhattacharya, Utso and Lewenstein, Maciej},
  journal = {Phys. Rev. Lett.},
  volume = {125},
  issue = {13},
  pages = {136801},
  numpages = {6},
  year = {2020},
  month = {Sep},
  publisher = {American Physical Society},
  doi = {10.1103/PhysRevLett.125.136801},
  url = {https://link.aps.org/doi/10.1103/PhysRevLett.125.136801}
}

@article{baldelli21,
  title = {Tracing non-Abelian anyons via impurity particles},
  author = {Baldelli, Niccol\`o and Juli\'a-D\'{\i}az, Bruno and Bhattacharya, Utso and Lewenstein, Maciej and Gra\ss{}, Tobias},
  journal = {Phys. Rev. B},
  volume = {104},
  issue = {3},
  pages = {035133},
  numpages = {10},
  year = {2021},
  month = {Jul},
  publisher = {American Physical Society},
  doi = {10.1103/PhysRevB.104.035133},
  url = {https://link.aps.org/doi/10.1103/PhysRevB.104.035133}
}

@article{bera24,
  title = {Steady-state quantum thermodynamics with synthetic negative temperatures},
  author = {Bera, Mohit Lal and Pandit, Tanmoy and Chatterjee, Kaustav and Singh, Varinder and Lewenstein, Maciej and Bhattacharya, Utso and Bera, Manabendra Nath},
  journal = {Phys. Rev. Res.},
  volume = {6},
  issue = {1},
  pages = {013318},
  numpages = {11},
  year = {2024},
  month = {Mar},
  publisher = {American Physical Society},
  doi = {10.1103/PhysRevResearch.6.013318},
  url = {https://link.aps.org/doi/10.1103/PhysRevResearch.6.013318}
}

@article{mohan25,
  title = {Coherent heat transfer leads to genuine quantum enhancement in the performances of continuous engines},
  author = {Mohan, Brij and Gangwar, Rajeev and Pandit, Tanmoy and Bera, Mohit Lal and Lewenstein, Maciej and Bera, Manabendra Nath},
  journal = {Phys. Rev. Appl.},
  volume = {23},
  issue = {4},
  pages = {044050},
  numpages = {25},
  year = {2025},
  month = {Apr},
  publisher = {American Physical Society},
  doi = {10.1103/PhysRevApplied.23.044050},
  url = {https://link.aps.org/doi/10.1103/PhysRevApplied.23.044050}
}

@Article{quspin1,
	title={{QuSpin: a Python package for dynamics and exact diagonalisation of quantum many body systems part I: spin chains}},
	author={Phillip Weinberg and Marin Bukov},
	journal={SciPost Phys.},
	volume={2},
	pages={003},
	year={2017},
	publisher={SciPost},
	doi={10.21468/SciPostPhys.2.1.003},
	url={https://scipost.org/10.21468/SciPostPhys.2.1.003},
}

@Article{quspin2,
	title={{QuSpin: a Python package for dynamics and exact diagonalisation of quantum many body systems. Part II: bosons, fermions and higher spins}},
	author={Phillip Weinberg and Marin Bukov},
	journal={SciPost Phys.},
	volume={7},
	pages={020},
	year={2019},
	publisher={SciPost},
	doi={10.21468/SciPostPhys.7.2.020},
	url={https://scipost.org/10.21468/SciPostPhys.7.2.020},
}

@software{code_ahm_hamiltonian,
  author       = {},
  title        = {sRb-root/Hybrid-Anyon-Otto-Thermal-machine-:
                   anyon\_engine
                  },
  month        = dec,
  year         = 2025,
  publisher    = {Zenodo},
  version      = {anyon\_engine\_latest},
  doi          = {10.5281/zenodo.17873351},
  url          = {https://doi.org/10.5281/zenodo.17873351},
}

@misc{Menon2025,
      title={Leveraging quantum statistics to enhance heat engines}, 
      author={Keerthy Menon and Thomas Busch and Thomás Fogarty},
      year={2025},
      eprint={2503.19341},
      archivePrefix={arXiv},
      primaryClass={quant-ph},
      url={https://arxiv.org/abs/2503.19341}, 
}

@article{Boubakour2023,
  title = {Interaction-enhanced quantum heat engine},
  author = {Boubakour, Mohamed and Fogarty, Thom\'as and Busch, Thomas},
  journal = {Phys. Rev. Res.},
  volume = {5},
  issue = {1},
  pages = {013088},
  numpages = {10},
  year = {2023},
  month = {Feb},
  publisher = {American Physical Society},
  doi = {10.1103/PhysRevResearch.5.013088},
  url = {https://link.aps.org/doi/10.1103/PhysRevResearch.5.013088}
}

@misc{supp,
    key = {Supplementary material available at link},
}

\clearpage
\onecolumngrid

\section*{Supplementary Material}

\section*{Derivation of the perturbative ground state energy $E_{per}$ in Eq.~9}
Let us define $\{\ket{C_\alpha}\}$ as the basis of Fock states, where $\ket{C_\alpha}\equiv\ket{n_{1,\alpha}n_{2,\alpha}\dots n_{j,\alpha}\dots n_{L,\alpha}}$ and $\hat{n}_j\ket{C_\alpha}=n_{j,\alpha}\ket{C_\alpha}$. For $U\ll J$, the first order perturbative correction to the ground state energy is given by the expectation value of the interaction term in the ground state $\ket{\psi_0}$ (corresponding to $U=0$), 
\begin{align}
E_{per}=\langle \psi_0|H_{int}|\psi_0\rangle&=\frac{U}{2}\sum_j\langle\psi_0|\hat{n}_j(\hat{n}_j-1)| \psi_0\rangle\nonumber\\
&=\frac{U}{2}\sum_j\sum_{\alpha,\beta}\langle\psi_0|C_\alpha\rangle\langle C_\alpha|\hat{n}_j(\hat{n}_j-1)|C_\beta\rangle\langle C_\beta| \psi_0\rangle\nonumber\\
&=\frac{U}{2}\sum_j\sum_{\alpha}|\langle C_\alpha|\psi_0\rangle|^2{n}_{j,\alpha}({n}_{j,\alpha}-1)\nonumber\\
&=\frac{U}{2}\sum_j\sum_{n=0}^Nn(n-1)\sum_{\alpha}|\langle C_\alpha|\psi_0\rangle|^2\delta_{n,n_{j,\alpha}}\nonumber\\
&=\frac{U}{2}\sum_j\sum_{n=0}^N n(n-1)P_j(n),
\end{align}
where $P_j(n)=\sum_{\alpha}|\langle C_\alpha|\psi_0\rangle|^2\delta_{n,n_{j,\alpha}}$ is the probability of the $j^{th}$ site of the lattice being occupied with $n$ particles in the ground state.

\begin{figure}[h]
    \centering
    \includegraphics[width=0.6\linewidth]{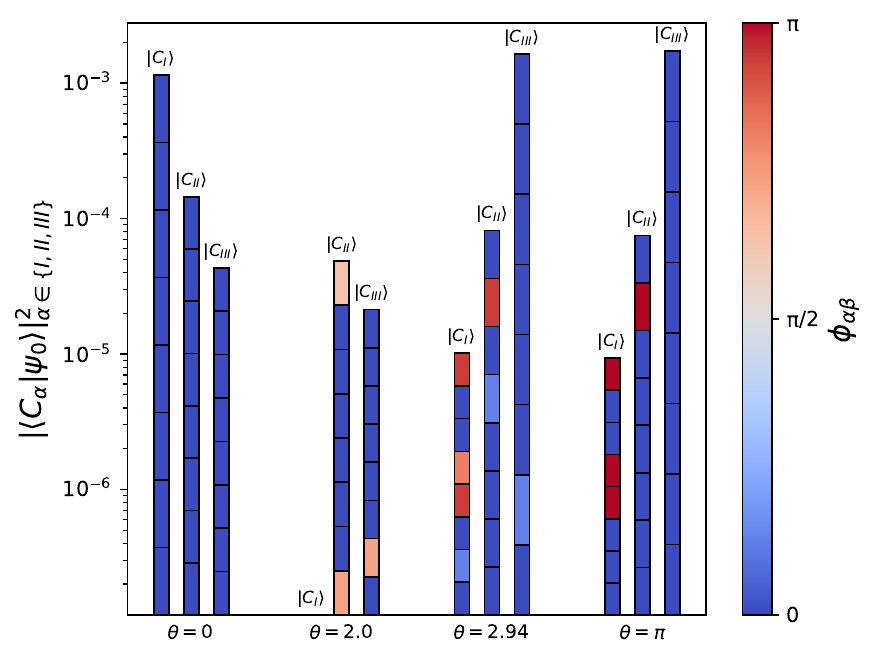}\label{fig:config_probs}
    \caption{Probability of finding the ground state $\ket{\psi_0}$ in the Fock-space configurations $\ket{C_{I, II, III}}$ (see main text) at different values of $\theta$. The blocks within each bar represent all the configurations in $\mathcal{S}_{\alpha=I, II, III}$, with the color of the blocks representing $\phi_{\alpha\beta}$. The numerical data for the plots are calculated for $L=12$, $N=6$, $J=1.0$ and $U=0$. }
\end{figure}

\section*{Additional details on the higher probability of multiple particle occupancy per site at $\theta\to0,\pi$  for $N\gtrsim L/2$}

For $\theta\neq0$, the presence of the density-dependent phase in the hopping term (Eq.~(1) of main text) implies a higher energy cost if multiple particles occupy the same site. For $N\gtrsim L/2$, the system cannot avoid such configurations resulting in a larger ground state energy except at $\theta\to 0, \pi $. To gain further intuition about the same, let us examine the probability of finding the ground state in a given Fock state $\ket{C_\alpha}$. Starting from the Schr\"{o}dinger equation for the ground state,  $H\ket{\psi_0}=E_0\ket{\psi_0}$, we find,
\begin{align}\label{eq:supp_prob}
|\langle C_\alpha|\psi_0\rangle|^2&=\frac{1}{E_0^2}|\langle C_\alpha|H|\psi_0\rangle|^2\nonumber\\
&=\frac{1}{E_0^2}\Big|\sum_\beta\langle C_\alpha|H|C_\beta\rangle\langle C_\beta|\psi_0\rangle\Big|^2\nonumber\\
&=\frac{1}{E_0^2}\Big[\sum_\beta|\langle C_\alpha|H|C_\beta\rangle|^2|\langle C_\beta|\psi_0\rangle|^2+\sum_{\substack{\beta,\gamma\\\gamma\neq\beta}}\langle C_\alpha|H|C_\beta\rangle\langle C_\gamma|H|C_\alpha\rangle\langle C_\beta|\psi_0\rangle\langle \psi_0|C_\gamma\rangle\Big]\nonumber\\
&=\frac{1}{E_0^2}\Big[\sum_\beta|\langle C_\alpha|H|C_\beta\rangle|^2|\langle C_\beta|\psi_0\rangle|^2+\sum_{\substack{\beta,\gamma\\\gamma<\beta}}\Big(\langle C_\alpha|H|C_\beta\rangle\langle C_\gamma|H|C_\alpha\rangle\langle C_\beta|\psi_0\rangle\langle \psi_0|C_\gamma\rangle+h.c.\Big)\Big]\nonumber\\
&=\frac{1}{E_0^2}\Big[\sum_\beta|\langle C_\alpha|H|C_\beta\rangle|^2|\langle C_\beta|\psi_0\rangle|^2+2\mathrm{Re}\sum_{\substack{\beta,\gamma\\\gamma<\beta}}\langle C_\alpha|H|C_\beta\rangle\langle C_\gamma|H|C_\alpha\rangle\langle C_\beta|\psi_0\rangle\langle \psi_0|C_\gamma\rangle\Big].
\end{align}
In the non-interacting limit $U=0$, we have $H=-J\sum_j(\hat{b}_j^\dagger e^{-i\hat{n}_j\theta}\hat{b}_{j+1}+\hat{b}_{j+1}^\dagger e^{i\hat{n}_j\theta}\hat{b}_{j})$. For each $\ket{C_\alpha}$, we define $S_\alpha$ as the set of Fock-states such that,  
$$\langle C_\alpha|H|C_\beta\rangle=\begin{cases}
-J_{\alpha\beta}e^{-i\phi_{\alpha\beta}} &\text{if }\ket{C_\beta}\in S_\alpha \\
0   & \text{if } \ket{C_\beta}\notin S_\alpha,
\end{cases}$$
where the values of $J_{\alpha\beta}$ and $\phi_{\alpha\beta}$ depend on the direction of hopping of the particle in $\langle C_\alpha|H|C_\beta\rangle$. In general, for $\ket{C_\alpha}=\ket{\dots n_{j,\alpha}n_{j+1,\alpha}\dots}$ and  $\ket{C_\beta}=\ket{\dots n_{j,\beta}n_{j+1,\beta}\dots}$, we have
\begin{equation}
J_{\alpha\beta}=J\sqrt{n_{j,\beta}(n_{j+1,\beta}+1)},\quad\phi_{\alpha\beta}=-n_{j,\alpha}\theta\quad {\rm for}~n_{j,\beta}-n_{j,\alpha}=n_{j+1,\alpha}-n_{j+1,\beta}=1\quad({\rm right~hopping})
\end{equation}
\begin{equation}
J_{\alpha\beta}=J\sqrt{n_{j+1,\beta}(n_{j,\beta}+1)},\quad\phi_{\alpha\beta}=n_{j,\beta}\theta\quad {\rm for}~n_{j,\beta}-n_{j,\alpha}=n_{j+1,\alpha}-n_{j+1,\beta}=-1\quad({\rm left~hopping})
\end{equation}
Substituting in Eq.~\eqref{eq:supp_prob}, we have
\begin{align}
|\langle C_\alpha|\psi_0\rangle|^2&=\frac{1}{E_0^2}\Big[\sum_{\ket{C_\beta}\in S_\alpha}J_{\alpha\beta}^2|\langle C_\beta|\psi_0\rangle|^2+2\mathrm{Re}\sum_{\substack{{\ket{C_\beta},\ket{C_\gamma}\in S_\alpha}\\\gamma<\beta}}J_{\alpha\beta}J_{\alpha\gamma}\langle C_\beta|\psi_0\rangle\langle \psi_0|C_\gamma\rangle e^{-i(\phi_{\alpha\beta}-\phi_{\alpha\gamma})}\Big].
\end{align}

The terms in the second summation in the above equation can destructively de-cohere for those $\ket{C_\alpha}$ which have high particle numbers per site, as for such states, $\phi_{\alpha\beta}-\phi_{\alpha\gamma}$ will fluctuate across the terms in the summation. It is therefore expected that such configurations will contribute to the ground state only if $\phi_{\alpha\beta}-\phi_{\alpha\gamma}=2m\pi$, for most of the configurations $\ket{C_\beta},\ket{C_\gamma}\in \mathcal{S}_\alpha$, where $m$ is an integer. For $\theta=0$, this is trivially satisfied and thus the probability for configurations with high occupancy per site is not reduced. However, in the limit $\theta\to\pi$, we have $\phi_{\alpha\beta}\to m\pi$ which allows for the existence of configurations for which $\phi_{\alpha\beta}-\phi_{\alpha\gamma}\to 2m\pi$. Thus, the probability of configurations with high occupancy per site is once again not prohibited in the limit $\theta\to\pi$, as in the case of $\theta=0$.

To illustrate the above, let us denote $\{\ket{C_{\alpha,3}}\}$ as the set of Fock-space configurations in which each of the configurations has at least one site with 3 particles. We examine the most-probable configurations $\ket{C_{I, II, III}}$ on this set, satisfying,  $|\langle C_I|\psi_0\rangle|^2_{\theta=0} =\max |\langle C_{\alpha,3}|\psi_0\rangle|^2_{\theta=0} $, $|\langle C_{II}|\psi_0\rangle|^2_{\theta=2.0} =\max |\langle C_{\alpha,3}|\psi_0\rangle|^2_{\theta=2.0} $, and $|\langle C_{III}|\psi_0\rangle|^2_{\theta=\pi} =\max |\langle C_{\alpha,3}|\psi_0\rangle|^2_{\theta=\pi} $. Note that there can be more than one configuration satisfying the conditions above; however, the following discussion doesn't depend on which of those configurations is chosen. Figure~\ref{fig:config_probs} shows the probabilities of finding the ground state in each of the above configurations at different values of $\theta$. The blocks within each bar represent the configurations in $\mathcal{S}_{\alpha=I, II, III}$, with the color of the blocks representing $\phi_{\alpha\beta}$. For $\theta=2.0$, it can be seen that even for the most probable configuration $\ket{C_{II}}$, there exist configurations  $\ket{C_\beta}$ in $\mathcal{S}_{\alpha=II}$ such $\phi_{\alpha\beta}\neq m\pi$. This results in a relatively smaller value of $|\langle C_{II}|\psi_0\rangle|^2$ as compared to the most probable configurations at $\theta=0,\pi$. As $\theta$ approaches the pseudo-fermionic limit (shown for $\theta=\pi-0.2$ and $\theta=\pi$ in Fig.~\ref{fig:config_probs}), the probability of the configuration $\ket{C_{III}}$ grows in magnitude as  $\phi_{\alpha\beta}\to 0,2\pi$ for all $\ket{c_\beta}$ in $\mathcal{S}_{\alpha=3}$.

\section*{Additional data for work output for different system sizes and filling fractions at low temperature for $U\neq 0$}

\begin{figure}[h]
    \centering
    \includegraphics[width=0.9\linewidth]{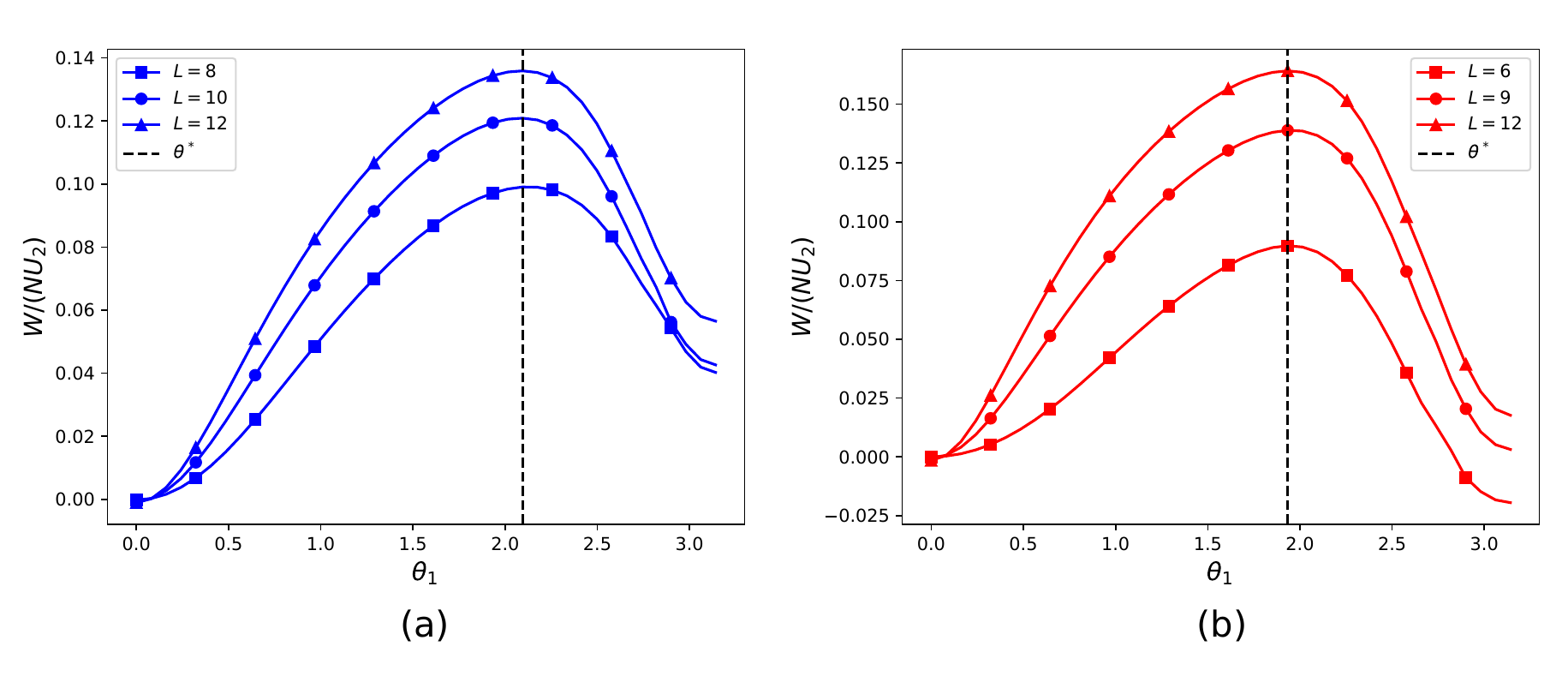}
    \caption{Low temperature ($T_A=T_B=0.1$) work output per particle (scaled with $U_2$) in the presence of finite interaction $U_2\neq0$ as a function of $\theta_1$ for different system sizes $L$ and filling fraction (a) $N/L=1/2$ and (b) $N/L=2/3$. The parameters chosen for simulation are the same as those used in Fig.~4a of main text, i.e., $U_1=0$, $J=1.0$ and $\theta_2=0$.}    \label{fig:supp_work_finiteU}

\end{figure}

In this section, we provide additional data to show how the finite work output at low temperature in the weakly interacting limit ($0<U\ll J$) scales with system size. From Fig.~\ref{fig:supp_work_finiteU}, we can see that the optimum value of $\theta_1$ for which the work is maximized does not appear to depend on the system size for a fixed filling. Furthermore, the maximum of the work is more pronounced for $N/L=2/3$ as compared to $N/L=1/2$. 

\end{document}